\documentclass[aps,twocolumn,pra,floatfix]{revtex4-1}
\usepackage{placeins}
\usepackage{amsmath,amssymb,bm}
\usepackage{graphicx}
\usepackage{epstopdf}
\usepackage{latexsym}
\usepackage{subfigure}
\usepackage[usenames,dvipsnames]{color}
\usepackage{hyperref}
\usepackage{natbib}
\usepackage{color}
\usepackage{lipsum}
\usepackage{textcomp}
  
\hypersetup{
  colorlinks,
  citecolor=Blue,
  linkcolor=Red,
  urlcolor=Blue}

\begin{document}
\newcommand{\cs}[1]{{\color{blue}$\clubsuit$#1}}

\title{Scattering of a dark-bright soliton by an impurity}
\author{Majed O. D. Alotaibi and Lincoln D. Carr}
\affiliation{Department of Physics, Colorado School of Mines, Golden, CO 80401, USA}
\begin{abstract}
We study the dynamics of a dark-bright soliton interacting with a fixed impurity using a mean-field approach. The system is described by a vector nonlinear Schrodinger equation (NLSE) appropriate to multicomponent Bose-Einstein condensates. We use the variational approximation, based on hyperbolic functions, where we have the center of mass of the two components to describe the propagation of the dark and bright components independently. Therefore, it allows the dark-bright soliton to oscillate. The fixed local impurity is modeled by a delta function. Also, we use perturbation methods to derive the equations of motion for the center of mass of the two components. The interaction of the dark-bright soliton with a delta function potential excites different modes in the system. The analytical model capture two of these modes: the relative oscillation between the two components and the oscillation in the widths. The numerical simulations show additional internal modes play an important role in the interaction problem. The excitation of internal modes corresponds to inelastic scattering. In addition, we calculate the maximum velocity for a dark-bright soliton and find it is limited to a value below the sound speed, depending on the relative number of atoms present in the bright soliton component and excavated by the dark soliton component, respectively. Above a critical value of the maximum velocity, the two components are no longer described by one center of mass variable and develop internal oscillations, eventually breaking apart when pushed to higher velocities.  This effect limits the incident kinetic energy in scattering studies and presents a smoking gun experimental signal.
\end{abstract} 
\maketitle    
\section{Introduction}     
Scattering is a fundamental physical process and essential tool to investigate objects in quantum theory~\cite{griffiths2005introduction}. We determine the low-energy interactions of subatomic particles by the well-known quantity, scattering length. Within this process, we acquire information regarding the nature of the interaction. Additionally, the interaction of solitons with localized impurities is a general and fundamental problem~\cite{Kevrekidis2008}. Utilizing the NLSE, many studies investigate the scattering of a bright or dark soliton with a localized impurity~\cite{Kivshar1989b,Kosevich1990,Lee2006,Frantzeskakis2002,Garnier2006,Seaman2005a,Sykes2009}. An impurity can be represented by a delta function as long as the size of the impurity is small enough compared to the soliton size. In BECs, one can create a delta function by a sharply focused far-detuned laser beam~\cite{Kevrekidis2008}. Of particular interest in soliton interactions with impurities is the interaction of two-component solitons with a delta function potential due to the rich dynamics that can be seen in these systems. The interaction of dark-bright solitons with an impurity has been the focus of other studies~\cite{Achilleos2011,Sykes2011}. But, to the best of our knowledge, the problem of the interaction of dark-bright solitons, with two independent centers of mass for the dark and bright components, with localized impurities and using the Lagrangian approach method has not been addressed so far. As we will show, the interplay between internal modes and the impurity is key to understanding the scattering process correctly. From an experimental point of view, one can interact with each component separately in multicomponent Bose-Einstein condensates~\cite{Hoefer2011b,Yan2012,Hamner2011b}. Thus it is possible to excite one component and not the other and achieve a dark-bright soliton with components of different centers of mass experimentally. 
   
In this work, we study the problem using coupled NLSEs, sometimes called the vector NLSE, that is appropriate to describe matter-wave dark-bright soliton in BEC~\cite{majed2017}. The existence of the delta function potential modifies the background of the dark soliton component, and therefore one should account for this effect. We do so by considering a perturbation method~\cite{Frantzeskakis2012,Kivshar1994} where we adjust the coupled NLSEs to account for the delta function as a small perturbation term. We proceed by adopting a modified Euler-Lagrange equation, called the variational Lagrangian approach, to calculate the equations of motion for the two propagating centers of mass (i.e., the locations of the dark component and the bright component)~\cite{Kivshar1995,Frantzeskakis2010a,Carretero-Gonzalez2008b}. The second part of this work is dedicated to investigating the dark soliton maximum velocity when interacting with a bright soliton in a dark-bright soliton. It is a well-known fact that the maximum velocity of a one-component dark soliton is the speed of sound~\cite{Kevrekidis2015}. We show that this qualitative characteristic of the dark soliton velocity is changing when we add a bright soliton to the picture. We adopt a known ansatz to describe the propagation of the dark-bright soliton. This ansatz is the exact solution for a dark-bright soliton with equally interacting coefficients (i.e., Manakov case~\cite{Salasnich2006}). We then extend our results numerically in the more general case. We show that the incident velocity and therefore kinetic energy of the dark-bright soliton on the impurity is limited by the number of atoms in the bright soliton relative to the ``hole'' or density notch made by the dark soliton.  Above a critical velocity, the dark-bright soliton develops oscillations, and when pushed further breaks up.  This sets definite limits on scattering studies. 
 
This article is organized as follows. In Sec.~\ref{sec:ANALYTICAL CALCULATIONS}, we study the scattering of the dark-bright soliton by a delta function potential using a variational approximation method based on a hyperbolic tangent (hyperbolic secant) for the dark (bright) soliton component for the two-component ansatz and utilizing a perturbation method to account for the effect of the delta function on the background. In Sec.~\ref{sec:VS_velocity}, we examine the velocity of the dark-bright soliton and obtain an analytical expression describing the effect of the bright component amplitude on the velocity of the dark-bright soliton. In Sec.~\ref{sec:FRHPRA:Numerical_Section_Scattering_of_dark-bright}, we investigate the scattering of the dark-bright soliton by a delta function potential by numerically integrating the dimensionless NLSE using an algorithm employing a pseudospectral method. We study the velocity of the dark-bright soliton numerically in Sec.~\ref{sec:FRHPRA:Numerical_Dark-bright soliton velocity}. Finally, we present our conclusions in Sec.~\ref{sec:FRHPRA:Conclusions}.
\section{ANALYTICAL CALCULATIONS}
\label{sec:ANALYTICAL CALCULATIONS}
%%%%%%%%%%%%%%%%%%%%%%%%%%%%%%%%%%%%%%%%%%%%%%%%%%%%%%%%%%%%%%%%%%%%%%%%%%%%%%%%%%%%%%%%%%%%%%%%
%%%%%%%%%%%%%%%%%%%%%%%%%    Lagrangian density and ansatz    %%%%%%%%%
%%%%%%%%%%%%%%%%%%%%%%%%%%%%%%%%%%%%%%%%%%%%%%%%%%%%%%%%%%%%%%%%%%%%%%%%%%%%%%%%%%%%%%%%%%%%%%%%
%
\subsection{Lagrangian density and ansatz}
\label{sec:Lagrangian density and ansatz}
\begin{figure}
	\centering
	\includegraphics[width=\columnwidth]{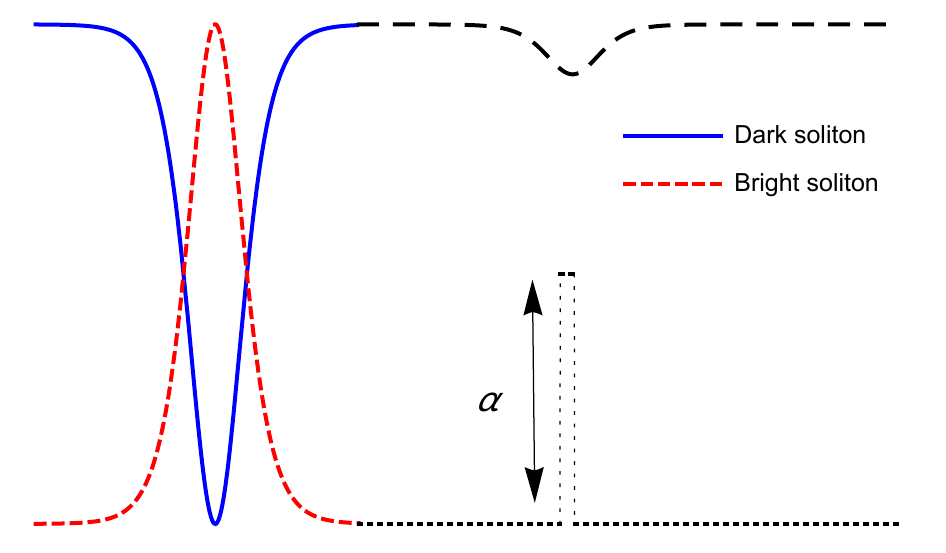}
	\caption{\emph{The effect of a delta function potential on the background of a dark-bright soliton. }. The delta function potential is modeled by  Thomas-Fermi cloud as described by Eq.~\eqref{eq:TF_scattering}. }
	\label{fig:FRHPRA:delta_function}
\end{figure}

%
%
%%%%%%%%%%%%%%%%%%%%%%%%%%%%%%%%%%%%%%%%%%%%%%%%%%%%%%%%%%%%%%%%%%%%%%%%%%%%%%%%%%%%%%%%%%%%%%%%%%%%%%%%%%
%%%%%%%%%%%%%%%%%%%%%%%%%%%%%%%%%%%%%%%%%%%%%%%%%%%%%%%%%%%%%%%%%%%%%%%%%%%%%%%%%%%%%%%%%%%%%%%%%%%%%%%%%%
%%%%%%%%%%%%%%%%%%%%%%%%%%%%%%%%%%%%%%%%%%%%%%%%%%%%%%%%%%%%%%%%%%%%%%%%%%%%%%%%%%%%%%%%%%%%%%%%%%%%%%%%%%
%
We start by introducing the coupled NLSEs:
\begin{align}
	\label{eq:coupled_NLSE}
&i \frac{\partial}{\partial t} u  + \frac{1}{2} \frac{\partial^2}{\partial x^2} u  - \left[g_{1}|u|^2 + g|v|^2 - u_{0}^2 \right] u = V(x) u, \\ \nonumber
&i \frac{\partial}{\partial t} v  +\frac{1}{2} \frac{\partial^2}{\partial x^2} v   - \left[g_{2}|v|^2 + g |u|^2  \right] v = V(x) v, 
\end{align}
where $u\equiv u(x,t)$ and $v\equiv v(x,t)$  are the wave functions for the dark and bright soliton components, respectively. The dark soliton wave function is rescaled to remove the background contribution, $u_{0}$, which is a standard procedure to avoid divergent normalization and energy~\cite{Kivshar1995}. The potential in the above equations takes the form,
\begin{align}
	\label{eq:delta_function}
 V(x) = \alpha \ \delta \left(x \right),  
\end{align}
for both components. We assume $\alpha \ll u_{0}$, and therefore we consider the potential to behave like a small perturbation effect which allows us to use the perturbation method. The same length-based units as we have described previously~\cite{majed2017} are used here: [x]=[L], [t]=[$L^2$], [$g_{1},g_{2},g$]=[$u_{0}$]=[$L^{-1}$], [$\alpha$]=[$\delta(x)$]=[$L^{-1}$], $|u,v|^2=[L^{-1}]$, where the square brackets mean ``the units of.'' The existence of a delta function affects the background of the dark-bright soliton, as seen in Fig.~\ref{fig:FRHPRA:delta_function},  and we need to modify the background also to account for this effect. We assume the dark soliton component lives on a modified Thomas--Fermi cloud, $|u_{\mathrm{TF}}|^2$, which accounts for the effect of the delta function on the background~\cite{Achilleos2011}, 
\begin{align}
	\label{eq:TF_scattering}
 |u_{\mathrm{TF}}|^2 \approx \frac{1}{g_{1}}(u^2_{0} - \alpha  u_{0} \mathrm{exp}(-2|x|)),
\end{align}
and by using the following transformations, 
\begin{equation}
	\label{eq:transformations}
\begin{aligned}
|u|^2 \rightarrow |u_{\mathrm{TF}}|^2 |u|^2 , \; |v|^2 \rightarrow \frac{|v|^2}{u^2_0}, \; t \rightarrow u^2_{0} t,  \; x \rightarrow u_{0} x,
\end{aligned}
\end{equation}
we recase Eqs.~\eqref{eq:coupled_NLSE} into the following:
\begin{align}
	\label{eq:coupled_NLSE_rescaled}
& i \frac{\partial }{\partial t}u +\frac{1}{2} \frac{\partial^2 }{\partial x^2}u - \left[g_{1} \left| u \right|^2 +g \left| v \right|^2 -1 \right] u = R_{u} \\ \nonumber
& i \frac{\partial }{\partial t}v +\frac{1}{2} \frac{\partial^2 }{\partial x^2}v - \left[g_{2} \left| v \right|^2  + g \left| u \right|^2  \right] v = R_{v}.
\end{align}
Where the RHS of Eqs.~\eqref{eq:coupled_NLSE_rescaled} represent the perturbation effects, 
\begin{align}
	\label{eq:RHS_scattering_potential}
& R_{u}=\frac{\alpha}{u_{0}} \left[(1-g_{1}|u|^2) u- \frac{x}{|x|} \frac{d}{dx} u  \right] e^{-2|x|} \\ \nonumber
& R_{v}=\frac{\alpha}{u_{0}} \left[  \delta \left(x\right)   - g |u|^2  e^{-2|x|}  \right] v, 
\end{align}
where $\alpha \ll 1$ in these units. We work with the following ansatz,
\begin{align}
	\label{eq:ansatz_scattering}
& u(x,t)= \frac{1}{\sqrt{g_{1}}} \{ c(t) \mathrm{tanh} \left[\frac{(x+d(t))}{\mathrm{w}(t)} \right] + i A(t) \} \\ \nonumber
& v(x,t)= \frac{1}{\sqrt{g_{2}}}  F(t) \mathrm{sech} \left[\frac{(x+b(t))}{\mathrm{w}(t)} \right] e^{i[\phi_{0}(t)+x \phi_{1}(t)]} .
\end{align}
Here $c(t)$ and $A(t)$ are the amplitude and velocity for the dark soliton component, respectively. The amplitude for the bright soliton component is $F(t)$. The velocity of the bright soliton is given by $\phi_{1}(t)$, and $d(t)$ and $b(t)$ are the position of the dark and bright soliton, respectively. The width for the two components is $\mathrm{w}(t)$ and $\phi_{0}(t)$ is a phase that gives rise to a complex amplitude of the bright component. We have a total of eight variational parameters that describe the propagation of the dark-bright soliton. The perturbation terms account for the effect of the potential (i.e., delta function). In the absence of the perturbation terms, the problem reduced to a propagation of the two--component dark-bright soliton~\cite{majed2017}. 
We use the normalization conditions,
\begin{subequations}
	\label{eq:normalizations}
	\begin{align}
	\int_{-\infty}^{\infty} dx\; \left(\frac{1}{g_{1}} - \left|u\right|^2\right)=\frac{N_{1}}{N}, \\  %=\frac{2 c^2 u^2_{0} w}{g_{1}}
	\int_{-\infty}^{\infty} dx\; \left| v\right|^2 =\frac{N_{2}}{N},										%= \frac{2 F^2 u^2_{0} w}{g_{2}}
\end{align}
\end{subequations}
for the dark and bright component, respectively. We subtract the background in the first component of Eqs.~(\ref{eq:normalizations}), therefore, $N_1$ is the number of atoms displaced by the dark soliton. We define the total number of atoms $N$ involved in the dark and bright solitons as
\begin{equation}
	\label{eq:total number of atmos}
	N_{1} + N_{2}=N,
\end{equation}
Using the ansatz, Eqs.~\eqref{eq:ansatz_scattering}, in the normalization, Eqs.~\eqref{eq:normalizations}, we find the relation between $N_{1}, N_{2}$ and the coefficients of the two components in the dark-bright soliton:
\begin{align}
\label{eq:relation1}
\frac{2 c^2 w}{g_{1}} = \frac{N_{1}}{N}, \\ \nonumber
\frac{2 F^2 w}{g_{2}} = \frac{N_{2}}{N}.
\end{align}
The modified Euler-Lagrange equation~\cite{Kivshar1995},
\begin{equation}
	\label{eq:modified_Euler_Lagrange_eq}
	\frac{\partial L}{\partial a_{j}} - \frac{d}{dt} \left(\frac{\partial L}{\partial a^{'}_{j}}  \right) = 2\ \text{Re} \left\{ \int_{-\infty}^{\infty}
	R^*_{u} \frac{\partial u}{\partial a_{j}} + R^*_{v} \frac{\partial v}{\partial a_{j}} \;dx   \right\}, 
\end{equation}
here $a_{j}$ represent the variational parameters in the ansatz. The Lagrangian density for the system of coupled equations, Eqs.~\eqref{eq:coupled_NLSE_rescaled} when $R_{u} = R_{v} =0$ is:
\begin{align}
	\label{eq:LagrangiandDensity}
\begin{split}
{\mathcal{L}} = & \frac{i}{2} \left[u^* \frac{\partial u}{\partial t} -u \frac{\partial u^*}{\partial t}  \right] \left[1-\frac{1}{g_{1}\left| u \right|^2} \right] - \frac{1}{2} \left|  \frac{\partial u }{\partial x}  \right|^2 \\ & 
- \frac{1}{2} \left[\sqrt{g_{1}} \left|u\right|^2 - \frac{1}{\sqrt{g_{1}}} \right]^2 +  \frac{i}{2} \left[v^* \frac{\partial v}{\partial t} -v \frac{\partial v^*}{\partial t}  \right]\\ &  - \frac{1}{2} \left|  \frac{\partial v }{\partial x}  \right|^2 
 -\frac{g_{2}}{2} \left| v \right|^4  - g \left| u \right|^2 \left| v \right|^2 .
\end{split}
\end{align}
We utilize the modified Euler-Lagrange equation, Eq.~\eqref{eq:modified_Euler_Lagrange_eq},  to account for the effect of the delta function on the background. By inserting the ansatz, Eq.~\eqref{eq:ansatz_scattering}, into the Lagrangian density, Eq.~\eqref{eq:LagrangiandDensity}, we obtain the Lagrangian as a function of the variational parameters. Then, we use Eq.~\eqref{eq:modified_Euler_Lagrange_eq} with the perturbation terms, Eq.~\eqref{eq:RHS_scattering_potential} to find the equations of motion (EOMs) of the system.
\subsection{Evolution equations}
\label{sec:Evolution equations}
The outcome of the calculations in Sec.~\ref{sec:Lagrangian density and ansatz} is a system of ordinary differential equations (ODEs) that describe the propagation of dark-bright solitons toward a delta function. Below, we write down only the equations that we are going to use to form a system of second order coupled ODEs,
\begin{subequations}
	\label{eq:EvolutionEqAll}
\begin{align}
%
%\phi_{1}
\label{eq:phi_evolution_eq}
&\frac{d}{dt} \phi_{1}(t) = \frac{g c^2(t)}{g_{1} \mathrm{w}^2(t)} \, \mathrm{csch}\left(\frac{b\left(t\right) - d\left(t\right) }{\mathrm{w}(t)}\right)^4 \\
\nonumber  & \times\left\{2  \left(b\left(t\right) - d\left(t\right)\right) \left[ 2 +\mathrm{cosh}\left( 2\frac{b\left(t\right) - d\left(t\right) }{\mathrm{w}(t)} \right) \right] \right.\\
\nonumber & \left.-3\mathrm{w}(t)\,\mathrm{sinh}\left( 2\frac{b\left(t\right) - d\left(t\right) }{\mathrm{w}(t)} \right) \right\} + \frac{g_{2} }{2 F^2(t) \mathrm{w}(t)} \Gamma_{1}, \\
% A
\label{eq:A_evolution_eq}
&\frac{d}{dt} A(t) = \frac{g c(t) F^2(t)}{2 g_{2} \mathrm{w}(t)} \, \mathrm{csch}\left(\frac{b\left(t\right) - d\left(t\right) }{\mathrm{w}(t)}\right)^4 \\
\nonumber  & \times\left\{2  \left(b\left(t\right) - d\left(t\right)\right) \left[ 2 +\mathrm{cosh}\left( 2\frac{b\left(t\right) - d\left(t\right) }{\mathrm{w}(t)} \right) \right] \right.\\
\nonumber & \left.-3\mathrm{w}(t)\,\mathrm{sinh}\left( 2\frac{b\left(t\right) - d\left(t\right) }{\mathrm{w}(t)} \right) \right\} - \frac{g_{1} }{4 c(t)} \Gamma_{2}, \\
% b
\label{eq:BS_evolution_eq}
&\frac{d}{dt} b\left(t\right) =  -\phi_{1}\left(t \right), \\
% d
\label{eq:DS_from_NLSE}
&\frac{d}{dt} d\left(t\right) =  - \frac{g_{1} N_{1} A(t)}{2 N \sqrt{1-A^2(t)}}.
\end{align}
\end{subequations}
Here, Eq.~\eqref{eq:DS_from_NLSE} is obtained by inserting the ansatz, Eq.~\eqref{eq:ansatz_scattering}, in the coupled NLSEs, Eqs.~\eqref{eq:coupled_NLSE_rescaled}, and separate the imaginary and real parts. In our calculations we take the delta function as located at the origin $x=0$ without loss of generality. We assume that the oscillations between the two component is very small (i.e. $b(t)-d(t) << 1$). 
The perturbation component $\Gamma_{1}$ in Eq.~\eqref{eq:phi_evolution_eq} is obtained by solving the RHS of Eq.~\eqref{eq:modified_Euler_Lagrange_eq} with $a_{j} = b$ and the perturbation component $\Gamma_{2}$ in Eq.~\eqref{eq:A_evolution_eq} is obtained by solving the RHS of Eq.~\eqref{eq:modified_Euler_Lagrange_eq} with $a_{j} = d$. As a result, we obtain the following terms,    
%
% The perturbation term, $\Gamma_{1} (\Gamma_{2})$, in equation~\eqref{eq:phi_evolution_eq} (~\eqref{eq:A_evolution_eq}) is obtained by solving the RHS of equation~\eqref{eq:modified_Euler_Lagrange_eq} with $a_{j} = b (a_{j} = d)$.
%
% Also, we used the normalization relation, equation~\eqref{eq:relation1}, in $\Gamma_{1}$ and $\Gamma_{2}$ and obtain the following terms,
\begin{subequations}
	\label{eq:gamma1}
\begin{align} 
% \Gamma_{1}
&\Gamma_{1} = \frac{8 g \alpha b(t) F^2(t) \exp(\frac{2 b(t)}{\mathrm{w}(t)})  }{g_{1} g_{2} u_{0} \mathrm{w}^2(t)} 
+ \frac{4 g \alpha c^2(t) F^2(t) }{15 g_{1} g_{2} u_{0}  \mathrm{w}^2(t)} \\\nonumber  & \times \left(b(t)-d(t)\right)  \mathrm{sech}^4\left(\frac{b(t)}{\mathrm{w}(t)}\right) \left\{-1+2 \mathrm{cosh} \left(\frac{2b(t)}{\mathrm{w}(t)}\right) \right\}  \\\nonumber  &
-\frac{8 g \alpha F^2(t)}{g_{1} g_{2} u_{0} \mathrm{w}(t)} \mathrm{sinh} \left(\frac{2b(t)}{\mathrm{w}(t)}\right) \mathrm{log}\left[1+\exp(\frac{2 b(t)}{\mathrm{w}(t)})   \right] \\\nonumber  &
+ \frac{2 g \alpha F^2(t)}{g_{1} g_{2} u_{0} \mathrm{w}(t)} \mathrm{cosh} \left(\frac{2b(t)}{\mathrm{w}(t)}\right) \mathrm{sech}^2 \left(\frac{b(t)}{\mathrm{w}(t)}\right)  \mathrm{tanh} \left(\frac{b(t)}{\mathrm{w}(t)}\right) \\\nonumber  &
- \frac{2 \alpha F^2(t) \left(-3 g + 3 g_{1} + g c^2(t)\right)}{3 g_{1} g_{2} u_{0} \mathrm{w}(t)}  \mathrm{sech}^2 \left(\frac{b(t)}{\mathrm{w}(t)}\right)  \mathrm{tanh} \left(\frac{b(t)}{\mathrm{w}(t)}\right)
\end{align}
% \end{subequations}
and for $\Gamma_{2}$ we get,
% \begin{subequations}
	\label{eq:gamma2}
\begin{align}
% \Gamma_{2}
&\Gamma_{2} =  \frac{\alpha c(t) \left[2+c^2(t) \mathrm{w}(t) \right]}{6 u_{0} \mathrm{w}^2(t)} \mathrm{sech}^2 \left(\frac{d(t)}{\mathrm{w}(t)}\right) \mathrm{tanh} \left(\frac{d(t)}{\mathrm{w}(t)}\right).
\end{align}
\end{subequations}
As a quick consistency check, note that when we set $\alpha =0$ (i.e. no potential), $\Gamma_{1}$ and $\Gamma_{2}$ are equal to zero too and therefore the perturbation terms are eliminated. By taking the second derivative of Eq.~\eqref{eq:BS_evolution_eq} and Eq.~\eqref{eq:DS_from_NLSE} we can further simplify the system of equations, Eqs.~\eqref{eq:EvolutionEqAll}, and obtain the following second order differential equations:
\begin{subequations}
	\label{eq:EvolutionEqAll_2nd_order}
\begin{align}
\label{eq:d_evolution_eq_2nd_order}
&\frac{d^2}{dt^2} d\left(t\right) = -\frac{g N_{2}}{4 N \mathrm{w}\left(t\right)} \, \mathrm{csch}\left(\frac{b\left(t\right) - d\left(t\right) }{\mathrm{w}\left(t\right)}\right)^4 \\
\nonumber  & \times\left\{2  \left(b\left(t\right) - d\left(t\right)\right) \left[ 2 +\mathrm{cosh}\left( 2\frac{b\left(t\right) - d\left(t\right) }{\mathrm{w}\left(t \right)} \right) \right] \right.\\
\nonumber & \left.-3\mathrm{w}\left(t\right)\,\mathrm{sinh}\left( 2\frac{b\left(t\right) - d\left(t\right) }{\mathrm{w}\left(t\right)} \right) \right\} + \frac{N \mathrm{w}^2\left(t\right) \Gamma_{2}}{2 N_{1}} , \\
% A
\label{eq:b_evolution_eq_2nd_order}
&\frac{d^2}{dt^2} b\left(t\right) = -\frac{g N_{1}}{2 N \mathrm{w}^3\left(t\right))} \, \mathrm{csch}\left(\frac{b\left(t\right) - d\left(t\right) }{\mathrm{w}\left(t\right)}\right)^4 \\
\nonumber  & \times\left\{2  \left(b\left(t\right) - d\left(t\right)\right) \left[ 2 +\mathrm{cosh}\left( 2\frac{b\left(t\right) - d\left(t\right) }{\mathrm{w}\left(t \right)} \right) \right] \right.\\
\nonumber & \left.-3\mathrm{w}\left(t\right)\,\mathrm{sinh}\left( 2\frac{b\left(t\right) - d\left(t\right) }{\mathrm{w}\left(t\right)} \right) \right\} + \frac{N  \Gamma_{1}}{ N_{2}},
\end{align}
\end{subequations}
%
%%%%%%%%%%%% check these %%%%%%%%%%%
%
where we used the normalization, Eq.~\eqref{eq:relation1}. Equations~\eqref{eq:EvolutionEqAll_2nd_order} describe the propagation of the two component dark-bright soliton in the vicinity of delta function potential located at $x=0$. 

\begin{figure}
  \centering
  \includegraphics[width=\columnwidth]{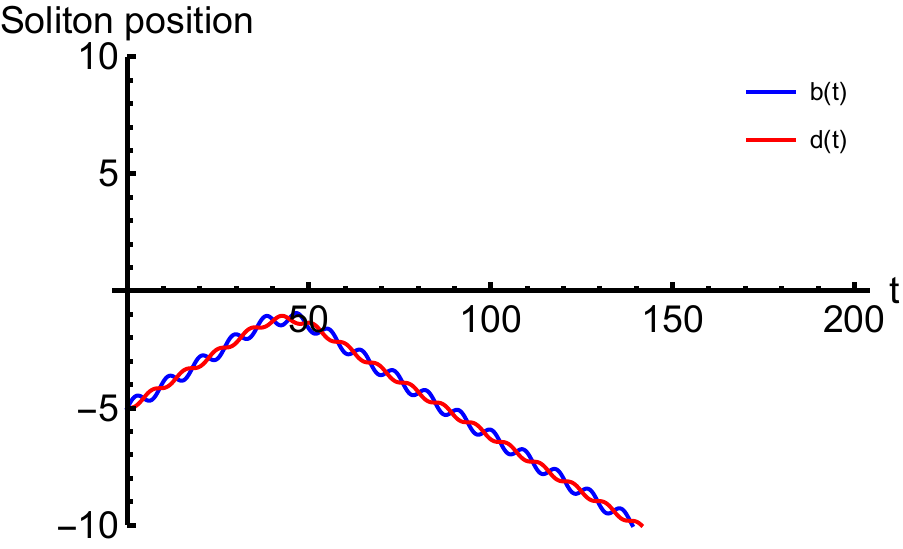}
  \caption{\emph{Reflection of dark-bright soliton.} We found the dark-bright soliton reflected by the potential when $\alpha = 0.15$. We set the center of mass velocity $V_{\mathrm{CM}} = 0.06$. }
  \label{fig:FRHPRA:scattering_1}
\end{figure}

\begin{figure}
  \centering
  \includegraphics[width=\columnwidth]{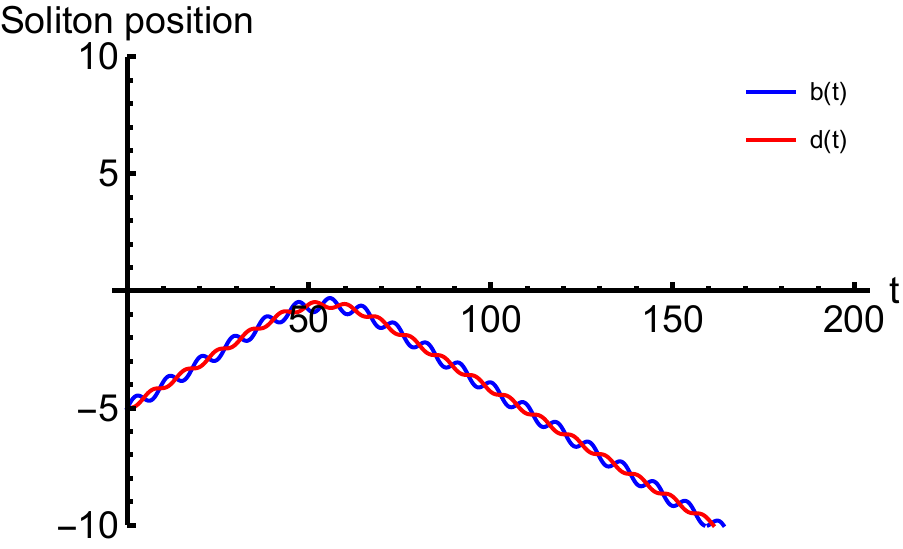}
  \caption{\emph{Reflection of dark-bright soliton with resonance.} Here we set $\alpha = 0.04$ and we see that the dark-bright soliton oscillates at the location of the potential for a finite time before it reflects back for the same value of $V_{\mathrm{CM}}$ used in Fig.~\ref{fig:FRHPRA:scattering_1}. Thus our model appears to capture a quasibound state or resonance.}
  \label{fig:FRHPRA:scattering_2}
\end{figure}

\begin{figure}
  \centering
  \includegraphics[width=\columnwidth]{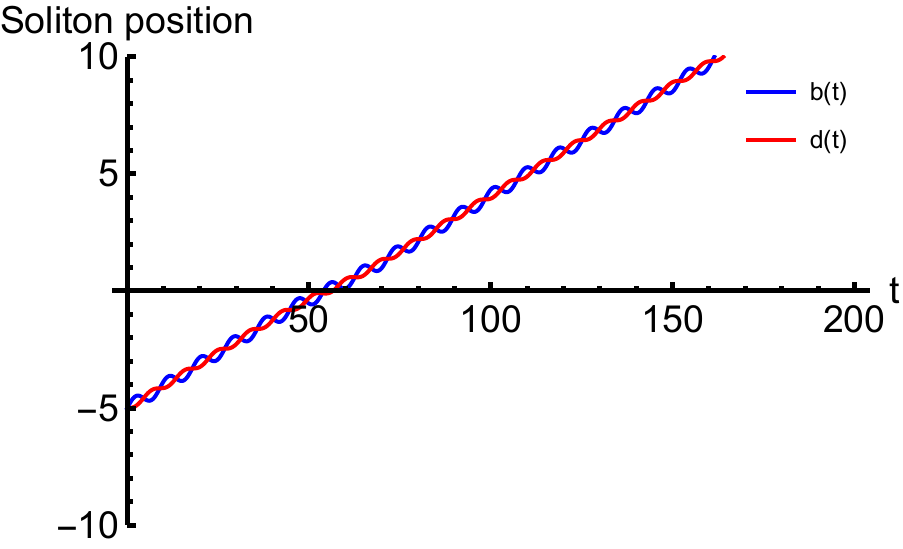}
  \caption{\emph{Transmission of dark-bright soliton.} Here we set $\alpha = 0.01$. The dark and bright soliton locations oscillate around their center of mass position. We found the dark-bright soliton passes over the potential for the same value of $V_{\mathrm{CM}}$ used in Fig.~\ref{fig:FRHPRA:scattering_1}.}
  \label{fig:FRHPRA:scattering_3}
\end{figure}

By fixing the initial velocity of the dark-bright soliton, $V_{\mathrm{CM}} = 0.06$ and depending on the strength of the potential, $\alpha$, we obtain three distinctive behavior of the dark-bright soliton as seen in Fig.~\ref{fig:FRHPRA:scattering_1}, \ref{fig:FRHPRA:scattering_2} and \ref{fig:FRHPRA:scattering_3}. These scenarios comprise reflection, reflection with resonance and a subsequent delay, and transmission, respectively.   
% Here, by varying the potential strength we capture three scenarios where we have the dark-bright soliton bouncing off the potential as seen in Fig.~\ref{fig:FRHPRA:scattering_1}. In Fig.~\ref{fig:FRHPRA:scattering_2} the dark-bright soliton hovering for a finite time over the potential barrier. Where as in the last figure Fig.~\ref{fig:FRHPRA:scattering_3}, we have a dark-bright soliton going through the potential.
In all figures, we find that the internal oscillation of the two components did not change for the incident and reflected dark-bright soliton. This means that there is ultimately no energy exchange between the internal modes and the kinetic energy of the dark-bright soliton. In Fig.~\ref{fig:FRHPRA:scattering_1}, \ref{fig:FRHPRA:scattering_2} and \ref{fig:FRHPRA:scattering_3} we set $g_{1}=2, g_{2}=2.7$, $g=2.6$, $\mathrm{w}=1$ and $N_{1}=0.521\times 10^5$. In Sec.~\ref{sec:FRHPRA:Numerical_Section_Scattering_of_dark-bright}, we compare these analytical predictions to the numerical calculations.  

%  
% \FloatBarrier
%
%%%%%%%%%%%%%%%%%%%%%%%%%%%%%%%%%%%%%%%%%%%%%%%%%%%%%%%%%%%%%%%%%%%%%%%%%%%%%%%%%%%%%%%%%%%%%%%%
%%%%%%%%%%%%%%%%%%%%%%%%%   	 Dark-bright soliton velocity   		 	   %%%%%%%%%%%%%%%%%
%%%%%%%%%%%%%%%%%%%%%%%%%%%%%%%%%%%%%%%%%%%%%%%%%%%%%%%%%%%%%%%%%%%%%%%%%%%%%%%%%%%%%%%%%%%%%%%%
%
\subsection{Dark-bright soliton velocity}
\label{sec:VS_velocity}  
\begin{figure}[!htbp]
\centerline{\includegraphics[width=\columnwidth]{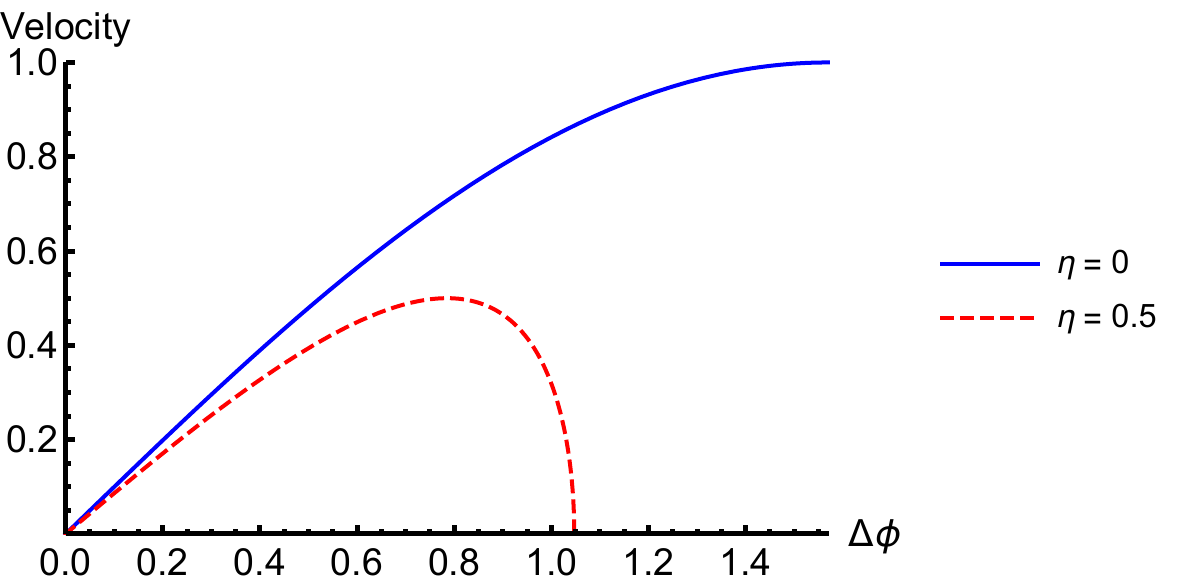}}
\caption{\label{fig:FRHPRA:fig1}\emph{Dark-bright soliton velocity}. The bright soliton amplitude is $\eta$. We set $\eta = 0.5$ in Eq.~\eqref{eq:EOM_velocity1_old} such that the amplitude of the bright component is half the amplitude of the dark component. Notice that the speed of sound, $c$, is 1 and the maximum velocity of the dark-bright soliton in this case is $c/2$ where we can calculate it from Eq.~\eqref{eq:EOM_velocity2}. Above $c/2$ the two components in the dark-bright soliton start to oscillate, as can be seen in the numerical simulation in Fig.~\ref{fig:FRHPRA:DB_velocity_num}, therefore an ansatz with one variable to describe the location of the two components is not valid. We plot the case for one-component dark soliton, $\eta=0$, for comparison.}
\end{figure}

In this section, we work with the velocity of the dark-bright soliton. Here we are working with different units~\cite{Achilleos2011}. The dimentionless coupled NLSEs,
\begin{align}
	\label{eq:coupledNLSE_VS}
i \frac{\partial \psi_{1}}{\partial t} &= -\frac{1}{2} \frac{\partial^2 \psi_{1}}{\partial x^2} + \left[\left| \psi_{1} \right|^2  +  \left| \psi_{2} \right|^2 -u^2_{0} \right] \psi_{1}, \nonumber \\
i \frac{\partial \psi_{2}}{\partial t} &= -\frac{1}{2} \frac{\partial^2 \psi_{2}}{\partial x^2} + \left[ \left| \psi_{2} \right|^2  +  \left| \psi_{1} \right|^2 \right] \psi_{2}.
\end{align}
are integrable (i.e., Manakov case) and possess an exact analytical dark-bright soliton solution of the following form~\cite{Alvarez2013a}:
\begin{align}
\label{eq:DB_solution}
% u(x,t)
\psi_{1}\left(x,t\right) &=   \mathrm{cos}\Delta \phi \; \mathrm{tanh} \xi + i \;\mathrm{sin}\Delta \phi ,   
\\ \nonumber  
% v(x,t)
\psi_{2}\left(x,t\right) &=   \eta \ \mathrm{sech} \xi \;  \mathrm{exp}{\left\{i \left[\phi_{0} + x \phi_{1} \right]\right\}}.
\end{align}
Here $\psi_{1}(x,t)$ and $\psi_{2}(x,t)$  are the wave functions for the dark and bright soliton components, respectively. The argument of the hyperbolic functions is $\xi=D\left(x-x_{0}\left(t\right)\right)$, $\mathrm{cos}\Delta \phi$ and $\eta$ are the dimensionless amplitudes of the dark and bright components, respectively, and D and $x_{0}\left(t\right)$ are the inverse width and the centre position of the dark-bright soliton. The phase jump over the dark soliton is $\Delta \phi$. 
By using the variational method, we obtain the EOMs,
\begin{align}
	\label{eq:EOM_velocity}
	\dot{x}_{0}&=D \; \mathrm{tan}\Delta \phi \\
	\label{eq:EOM_width}
	D^2&= \mathrm{cos}^2\Delta \phi - \eta^2
\end{align}
Plugging Eq.~\eqref{eq:EOM_width} into Eq.~\eqref{eq:EOM_velocity}, we get:
\begin{align}
	\label{eq:EOM_velocity1_old}
	 \dot{x}_{0} &= \sqrt{ \mathrm{cos}^2\Delta \phi -  \eta^2} \; \mathrm{tan}\Delta \phi. 
\end{align}
For $\eta=0$ (i.e. $v(x,t)=0$), we have $\dot{x_{0}}=\mathrm{sin}\Delta \phi$ which is the velocity of dark soliton in one-component BECs, a Josephson-type relation based on the phase jump phi over the soliton~\cite{Reinhardt1997b}. The two extreme limits of the phase jump over the dark soliton are $\Delta \phi=0$ and $\Delta \phi=\frac{\pi}{2}$. In the former the depth of the dark soliton is maximum, and the velocity is zero. In the latter case, the depth of the dark soliton is zero whereas the velocity is maximum (i.e., the speed of sound, $c$). By examining Eq.~\eqref{eq:EOM_velocity1_old}, we find that the existence of a bright component affects the velocity of the dark-bright soliton and sets an upper limit for the maximum velocity depending on the amplitude of the bright component. Also, the term $\mathrm{cos}^2\Delta \phi -  \eta^2 $ in Eq.~\eqref{eq:EOM_velocity1_old}  restricts the range of the real values of the velocity of the dark-bright soliton. By equating this term to zero, we find that  $\Delta \phi$ gives a real value only for $\Delta \phi:0 \rightarrow \mathrm{cos}^{-1}\left( \eta \right)$. This implies that there is  a finite range of the velocity of the dark-bright soliton as well as a finite range of the depth of the dark component in the dark-bright soliton. Since the depth of the dark component is governed by $\mathrm{cos}\Delta \phi$, the range of the dark soliton amplitude goes from $u_{0}$ when $\Delta \phi=0$ to $\eta$ when $\Delta \phi= \mathrm{cos}^{-1}\left( \eta \right)$. That is, the minimum depth of the dark soliton component in the dark-bright soliton is not zero as it is the case for one-component dark soliton. It depends on the amplitude of the bright soliton component. In the range $\Delta \phi:0 \rightarrow \mathrm{cos}^{-1}\left( \eta\right)$ the dark-bright soliton velocity is zero on both ends as seen form Eq.~\eqref{eq:EOM_velocity1_old}. So, in this interval, the velocity increases to a finite value and decreases, see Fig.~\ref{fig:FRHPRA:fig1}. To find the maximum velocity of the dark-bright soliton we differentiate Eq.~\eqref{eq:EOM_velocity1_old} and solve it for $\Delta \phi$. As a result, we obtain the following equation,
\begin{align}
	\label{eq:EOM_velocity2}
	\dot{x}_0^{\mathrm{max}} &= 1- \eta = 1 - \sqrt{\frac{N_{2} D}{2 (N_{1}+N_{2})}}.
\end{align}
Above this maximum value, $\dot{x}_0^{\mathrm{max}}$, an internal oscillation develops between the two components which means the two component are no longer described by one center of mass variable for the dark-bright soliton. Therefore, the above ansatz, Eq.~\eqref{eq:DB_solution}, is not valid beyond this maximum velocity. Note that for $\eta \rightarrow 1$, $\dot{x}_0^{\mathrm{max}} \rightarrow 0$, $N_{2} \rightarrow N_{1}$ from below, and the dark-bright soliton ceases to exist, as shown in~\cite{majed2017}. In Fig.~\ref{fig:FRHPRA:fig1} we plot the velocity of the dark-bright soliton for $\eta = 0.5$ (i.e., the bright component is half the amplitude of the dark component). Since the amplitude squared of both components is proportional to the number of atoms in each component, the case where $\eta = 0.5$ is equivalent to $N_{1}=2 N_{2}$, where $N_{1}$ is the number of atoms displaced by the dark soliton and $N_{2}$ is the number of atoms in the bright soliton, as described in Sec.~\ref{sec:Lagrangian density and ansatz}. 
%%%%%%%%%%%%%%%%%%%%%%%%%%%%%%%%%%%%%%%%%%%%%%%%%%%%%%%%%%%%%%%%%%%%%%%%%%%%%%%%%%%%%%%%%%%%%%%%
%%%%%%%%%%%%%%%%%%%%%%%%%%%%%%%%          	 Numerical Section          %%%%%%%%%%%%%%%%%%%%%%%%
%%%%%%%%%%%%%%%%%%%%%%%%%%%%%%%%%%%%%%%%%%%%%%%%%%%%%%%%%%%%%%%%%%%%%%%%%%%%%%%%%%%%%%%%%%%%%%%%
%
% \FloatBarrier
\section{NUMERICAL CALCULATIONS}
	\label{sec:FRHPRA:Numerical_Section}
We numerically study the interaction between the two components in the dark-bright soliton and the potential barrier in Sec.~\ref{sec:FRHPRA:Numerical_Section_Scattering_of_dark-bright} where we use a delta function as described by Eq.~\eqref{eq:delta_function}. The strength of the delta function potential can be modified by varying the amplitude $\alpha$. In addition, we study the effect on a one-component dark soliton velocity when interacting with another component, in this case a one-component bright soliton. The velocity of the dark soliton component is fundamentally different than the velocity of the bright soliton. As we increase the speed of the dark soliton, its width goes to infinity, and the depth goes to zero. As a result, the dark soliton disappears and we left with a plane wave. Also, the maximum velocity of a one-component dark soliton is the speed of sound in BEC. In contrast, the one-component bright soliton velocity is unbounded and its width is not a function of its velocity at all. These known facts raise questions when we are dealing with the dark-bright soliton, as we explored under certain simplifying assumptions amenable to analytical treatment in Sec.~\ref{sec:VS_velocity}, where we found a maximum velocity dependent on the difference between the amplitudes of the two components. Therefore, the presence of the bright soliton component will qualitatively change the behavior of the well-known dark soliton velocity. We now relax those assumptions to treat the general case numerically. Throughout this section, we performed the simulations with grid size $n_{x} = 256$ in a box with hard-wall boundaries. The box length was set to $L = 100$ unless otherwise noted.
\subsection{Scattering of dark-bright soliton by potential barrier}
\label{sec:FRHPRA:Numerical_Section_Scattering_of_dark-bright}
We now explore the scattering problem numerically by creating a moving dark-bright soliton incident on a delta-function potential.  We make no other assumptions, allowing for internal excitations of the dark-bright soliton around its center of mass. There are two ways to to impart a velocity to the dark-bright soliton. The first is to imprint a linear phase ramp on the bright soliton component.  As a result the bright soliton will drag the dark soliton, and therefore we will have a moving dark-bright soliton. 

The second is to imprint a phase to one side of a dark soliton, creating a phase jump $\Delta \phi$, therefore, we obtain the same moving dark-bright soliton. There is however a significant difference in the outcome in terms of excitation of internal modes. In the first case, imprinting a phase on the bright soliton will produce an internal oscillation of the two components of the dark-bright soliton. We use this method here to move the dark-bright soliton. The second method is used in the second part of the numerical section where we are interested in having the two components move without any internal oscillation.

\begin{figure}[!htbp]
\begin{subfigure}
  \centering
  \includegraphics[width=0.95\columnwidth]{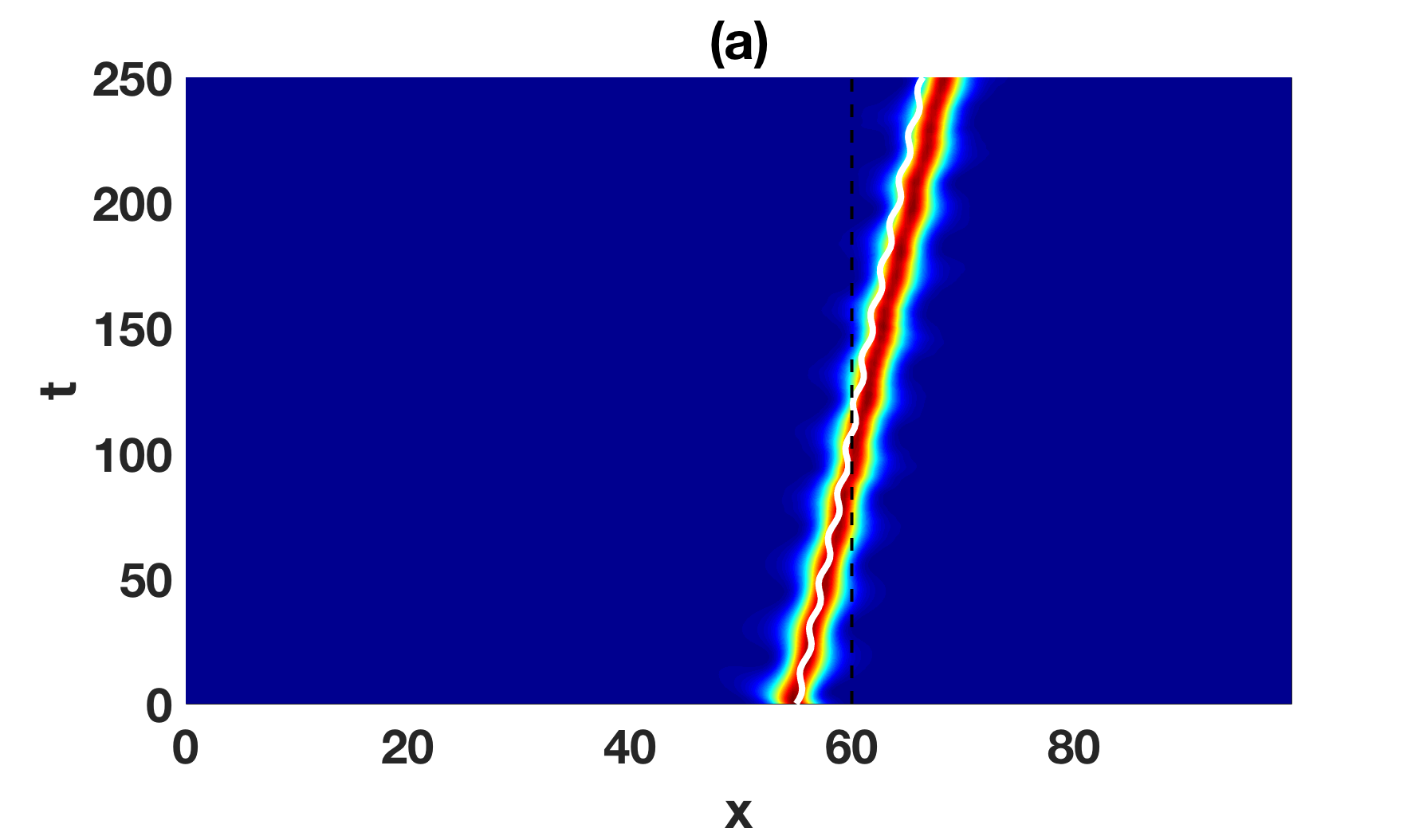}
  %\caption{1a}
  %\label{fig:phi_2}
\end{subfigure}%
\begin{subfigure}
  \centering
  \includegraphics[width=0.95\columnwidth]{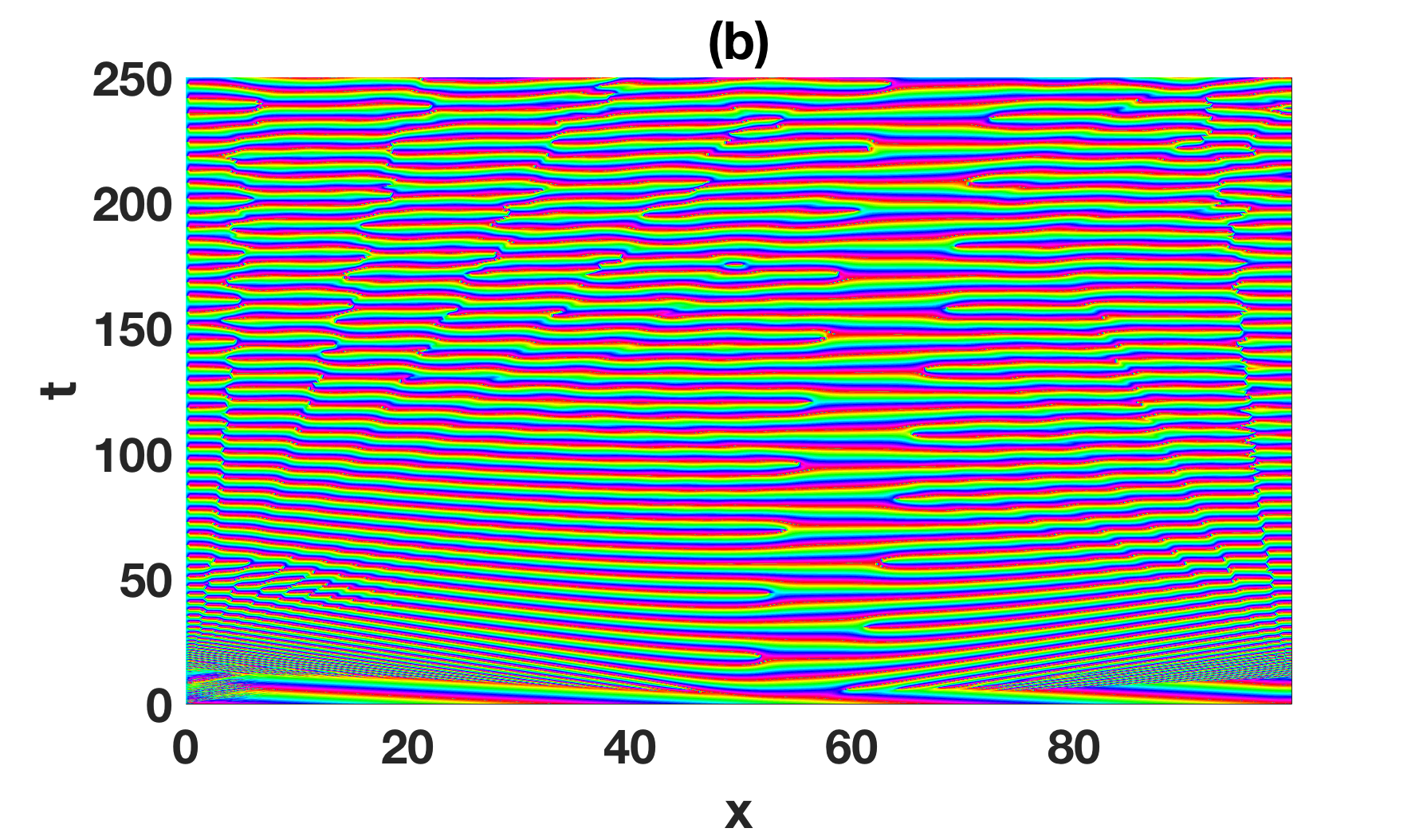}
  %\caption{1b}
  %\label{fig:phi_6}
\end{subfigure}%
\begin{subfigure}
  \centering
  \includegraphics[width=0.95\columnwidth]{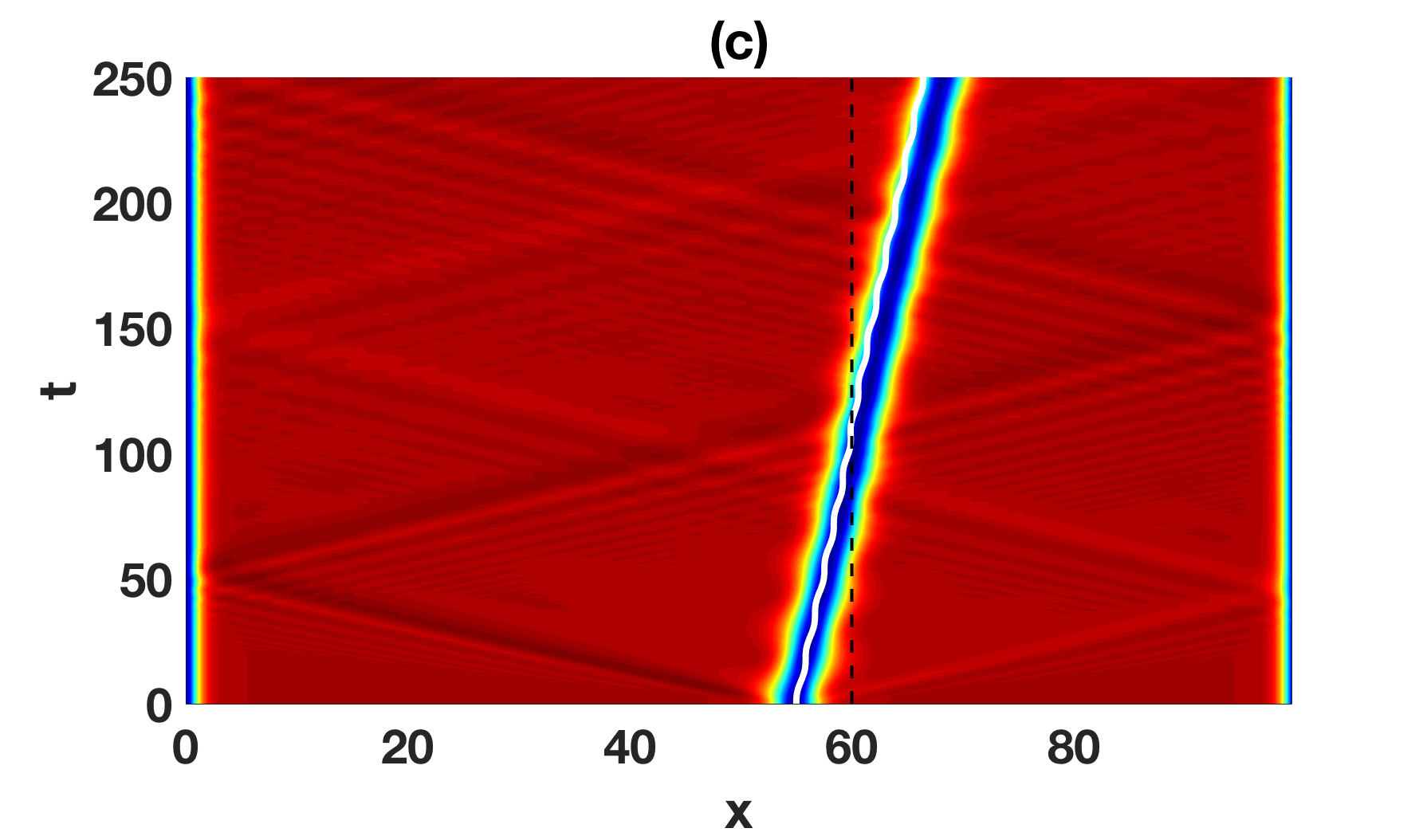}
  %\caption{1b}
  %\label{fig:phi_6}
\end{subfigure}
\begin{subfigure}
  \centering
  \includegraphics[width=0.95\columnwidth]{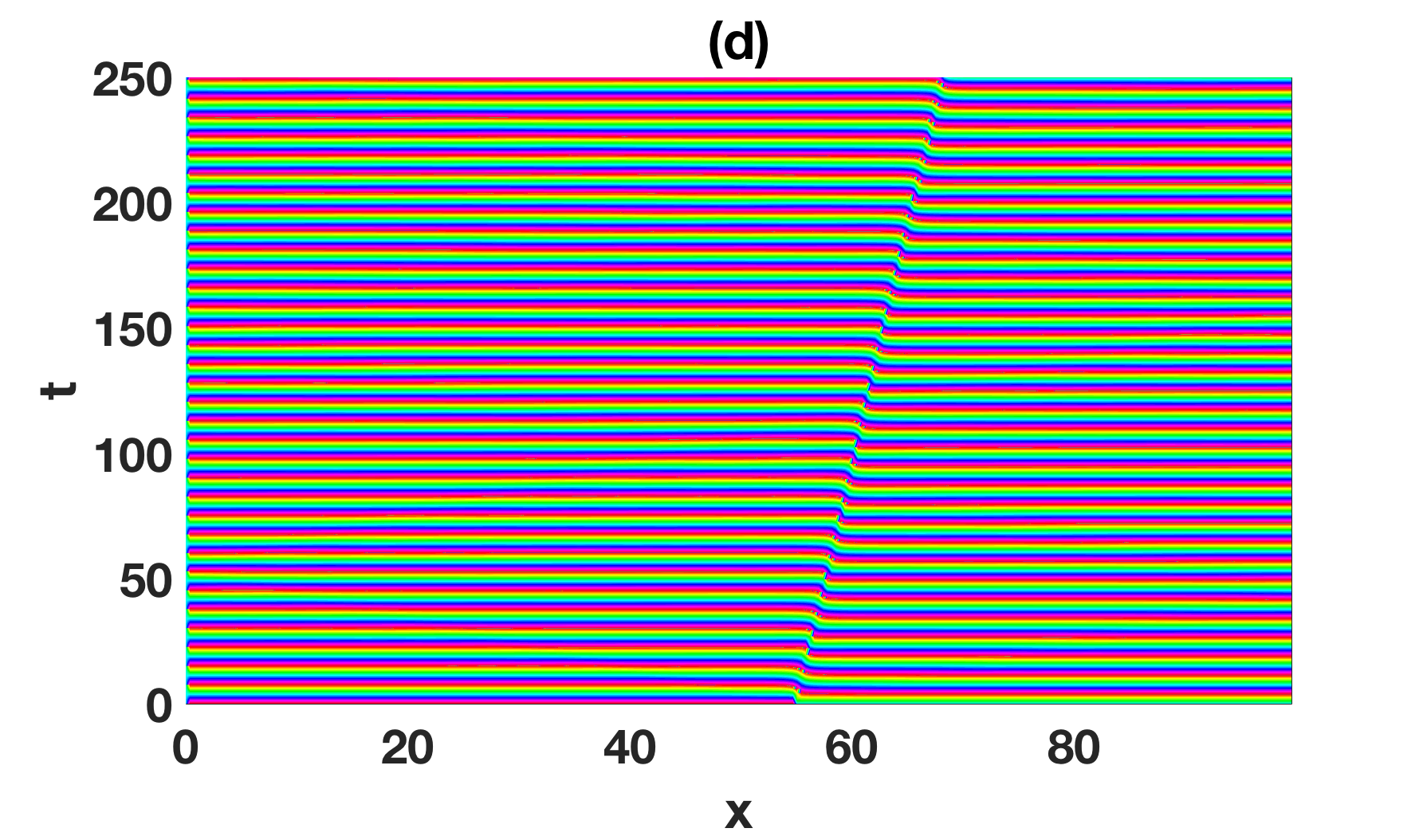}
  %\caption{1b}
  %\label{fig:phi_6}
\end{subfigure}
\caption{\emph{Transmission of a dark-bright soliton.} (a) Density and (b) phase of the bright soliton; (c) density and (d) phase of the dark soliton. The kinetic energy of the two-component dark-bright soliton is greater than the potential energy of the barrier and therefore the dark-bright soliton passes over it. The phonons appear as bright yellow bands moving at a much higher velocity, primarily associated with the initial velocity kick applied at $t=0$. We set  $\alpha = 0.01$ and $V_{\mathrm{CM}} = 0.06$. The delta function located at $x=60$. The white thick line represents the analytical results.}  
  \label{fig:FRHPRA:all_fig_delta_0-01}
\end{figure}

We thus first imprint a phase on the bright component and therefore the dark-bright soliton moves toward the delta function which is for convenience located at $x=60$ in our simulation, with the grid of $256$ points running from $x=0$ to $x=100$. Depending on the strength of the delta function (i.e., $\alpha$), where we fixed the incident velocity for all cases,  we have three distinctive sets of dark-bright soliton dynamics ensue. In Fig.~\ref{fig:FRHPRA:all_fig_delta_0-01}, where we have both the analytical and numerical results plotted on the same graph,  we set $\alpha=0.01$ and find that the dark-bright soliton is passing over the potential. When the dark-bright soliton interacts with the delta function, we found that numerically the dark-bright soliton moves slightly faster than the analytical prediction. At the end of this section, we discuss the physical reasons for the discrepancy between the analytical and numerical results. 

\begin{figure}[!htbp]
\begin{subfigure}
  \centering
  \includegraphics[width=0.95\columnwidth]{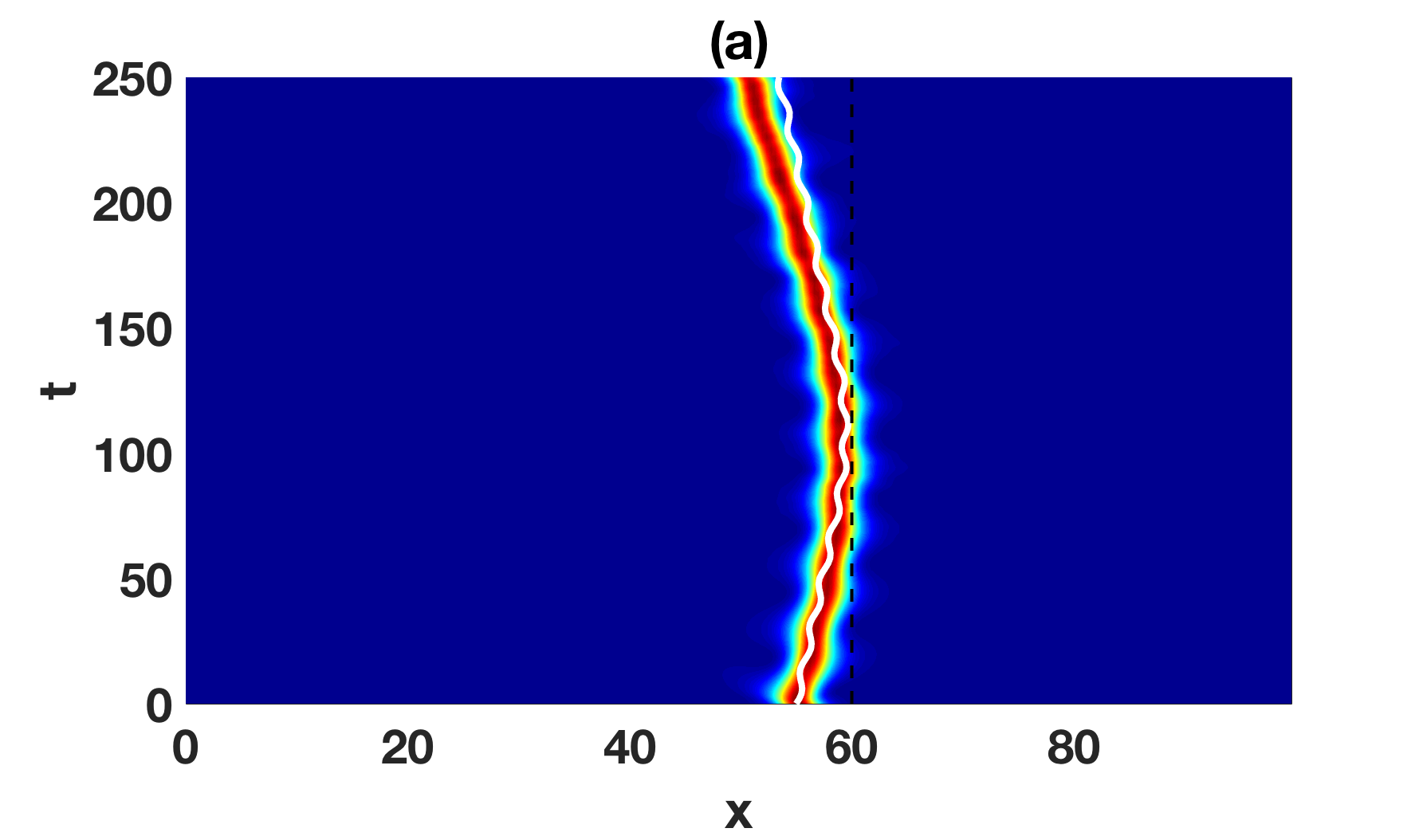}
  %\caption{1a}
  %\label{fig:phi_2}
\end{subfigure}%
\begin{subfigure}
  \centering
  \includegraphics[width=0.95\columnwidth]{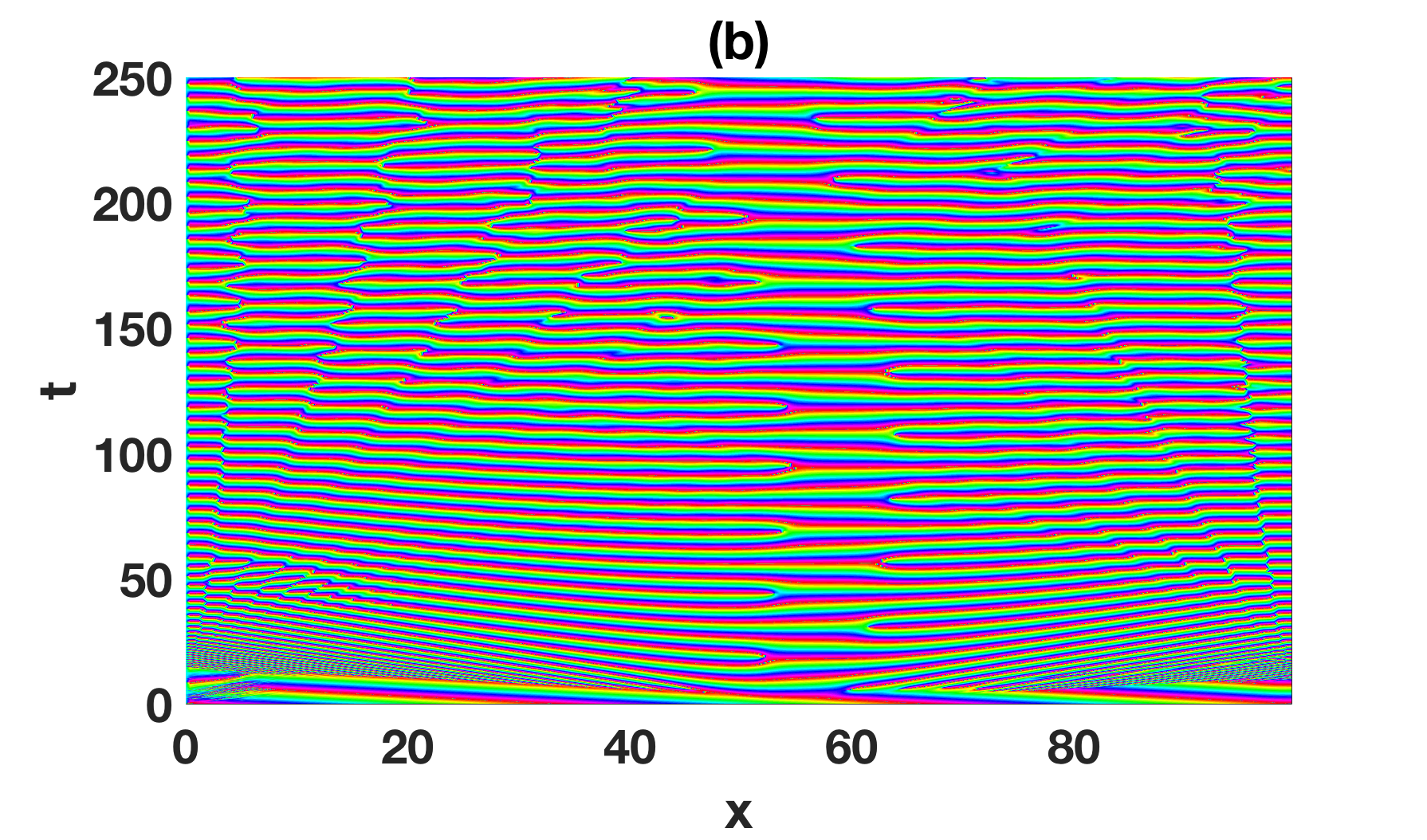}
  %\caption{1b}
  %\label{fig:phi_6}
\end{subfigure}%
\begin{subfigure}
  \centering
  \includegraphics[width=0.95\columnwidth]{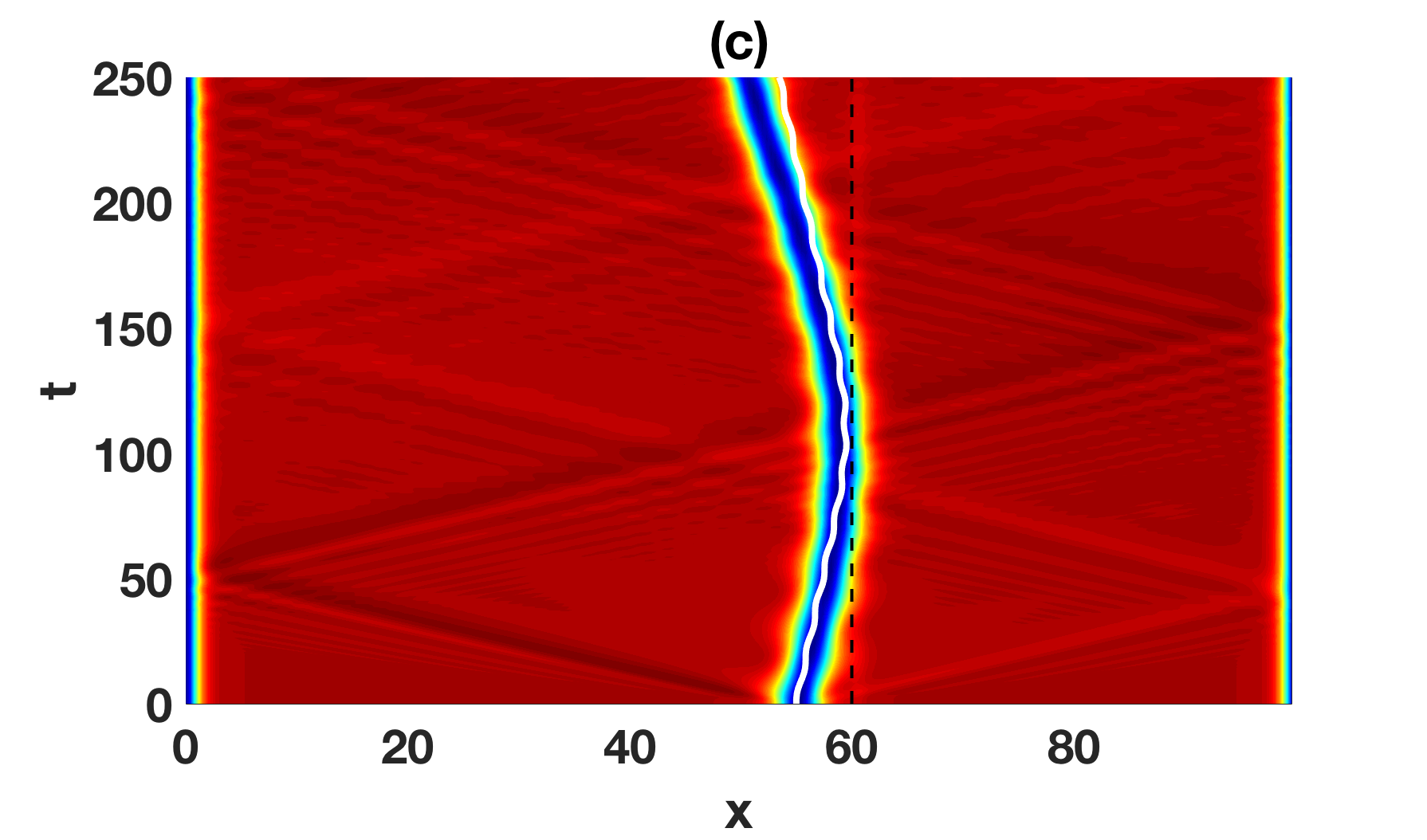}
  %\caption{1b}
  %\label{fig:phi_6}
\end{subfigure} 
\begin{subfigure}
  \centering
  \includegraphics[width=0.95\columnwidth]{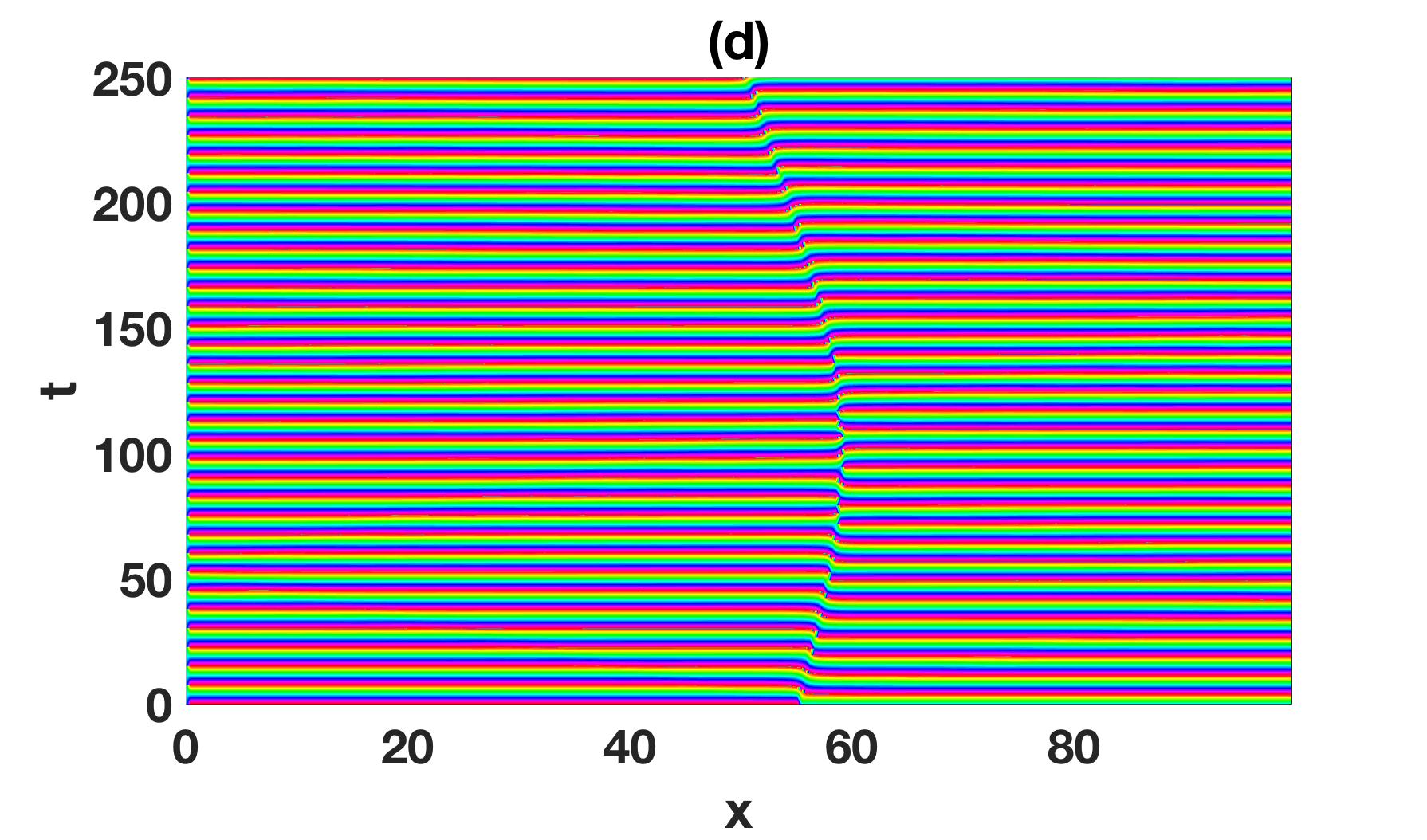}
  %\caption{1b}
  %\label{fig:phi_6}
\end{subfigure}
\caption{\emph{Resonant reflection of a dark-bright soliton.} (a) Density and (b) phase of the bright soliton; (c) density and (d) phase of the dark soliton. The kinetic energy of the two components dark-bright soliton is almost equal to the potential energy of the barrier and therefore the dark-bright soliton hovers over the barrier for a finite time where energy goes into internal modes, not phonons. We set  $\alpha = 0.04$ and $V_{\mathrm{CM}} = 0.06$. The delta function located at $x=60$. The white thick line represents the analytical results.} 
  \label{fig:FRHPRA:all_fig_delta_0-04} 
\end{figure}

In Fig.~\ref{fig:FRHPRA:all_fig_delta_0-04}, we set $\alpha= 0.04$, and the outcome of this comparison between the analytical and numerical calculations is that the dark-bright soliton hovers around the location of the potential for a finite time, appearing to be briefly quasibound or resonant, and then is reflected. The analytical predictions and the numerical calculations show that the dark-bright soliton reflects with different velocities. We consider this case as an inelastic scattering of the dark-bright soliton by a delta function as can be seen in Fig.~\ref{fig:FRHPRA:summary_plot}. Numerically, when the dark-bright soliton interacts with the potential barrier an internal state is excited (i.e., the internal oscillation of the two components) and therefore the dark-bright soliton come out of the interaction with a different velocity than the initial one.

\begin{figure}[!htbp]
\begin{subfigure}
  \centering
  \includegraphics[width=0.95\columnwidth]{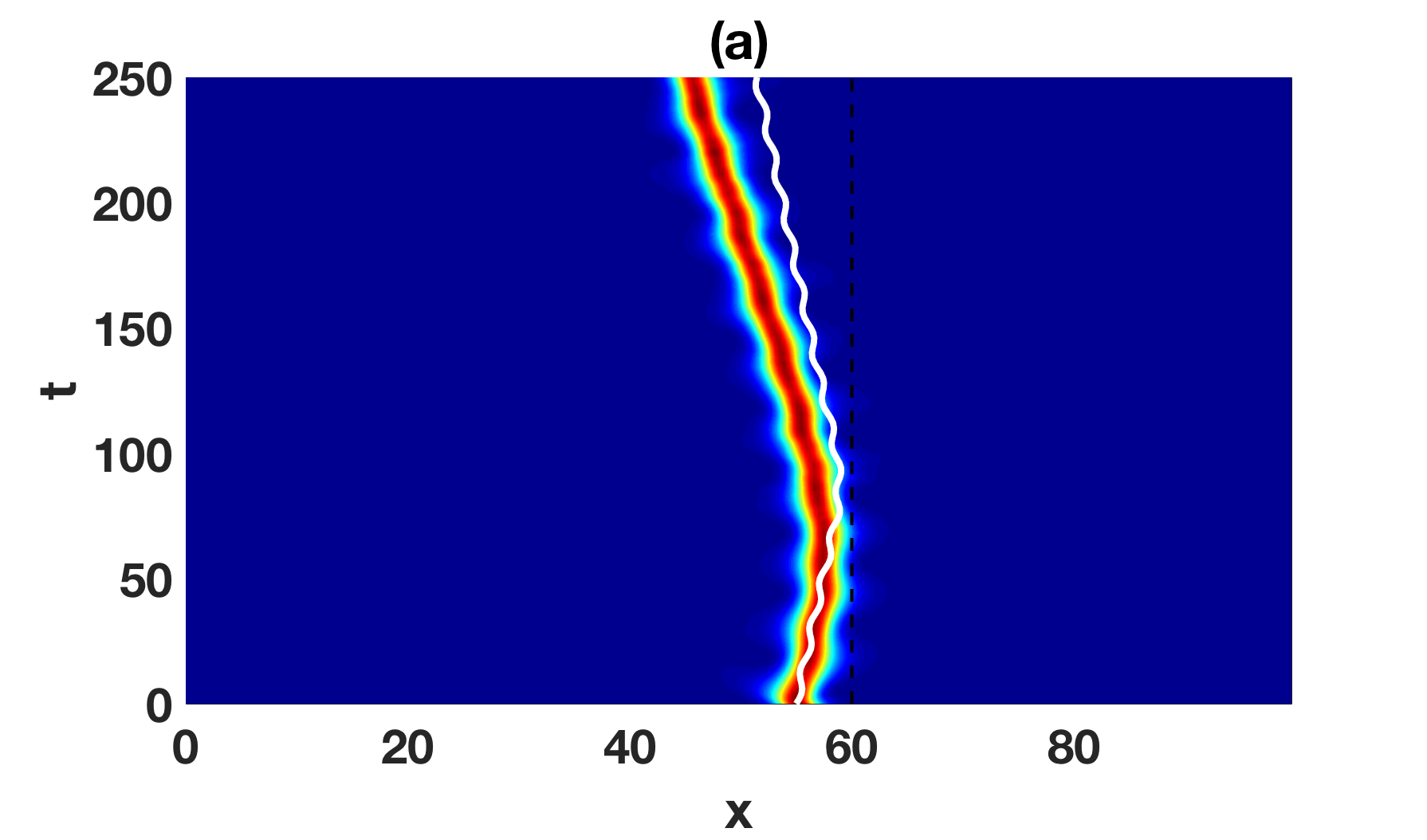}
  %\caption{1a}
  %\label{fig:phi_2}
\end{subfigure}%
\begin{subfigure}
  \centering
  \includegraphics[width=0.95\columnwidth]{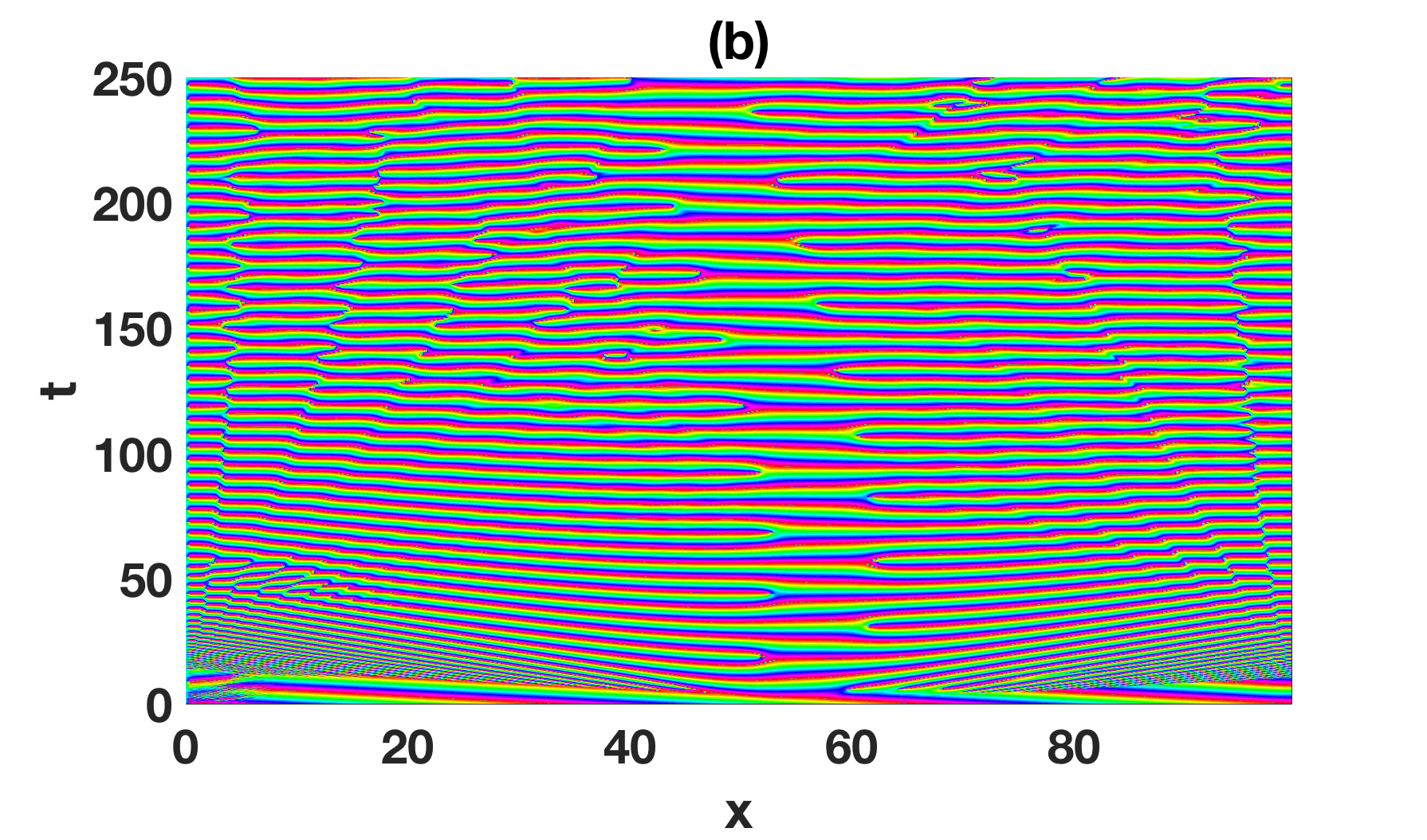}
  %\caption{1b}
  %\label{fig:phi_6}
\end{subfigure}%
\begin{subfigure}
  \centering
  \includegraphics[width=0.95\columnwidth]{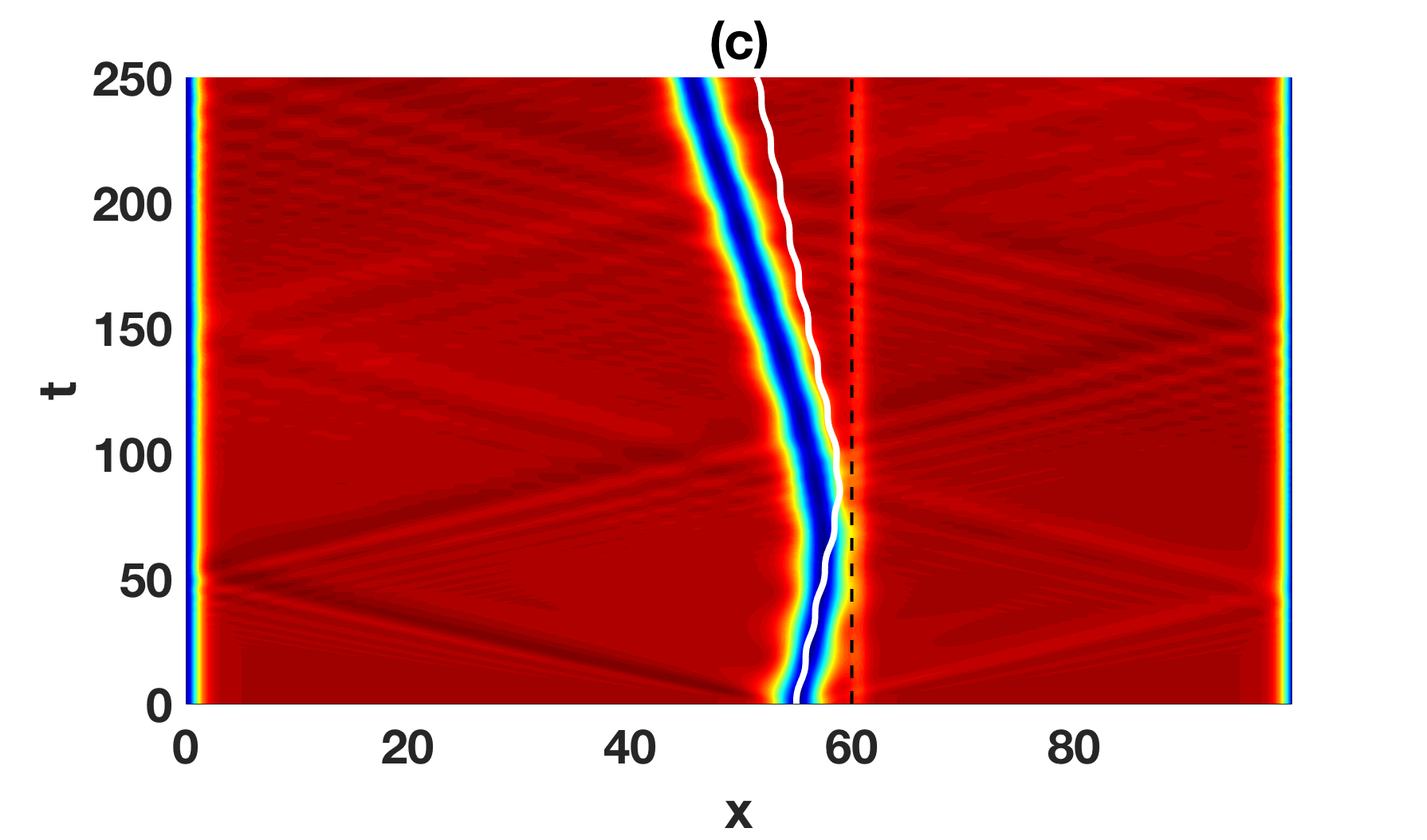}
  %\caption{1b}
  %\label{fig:phi_6}
\end{subfigure}
\begin{subfigure}
  \centering
  \includegraphics[width=0.95\columnwidth]{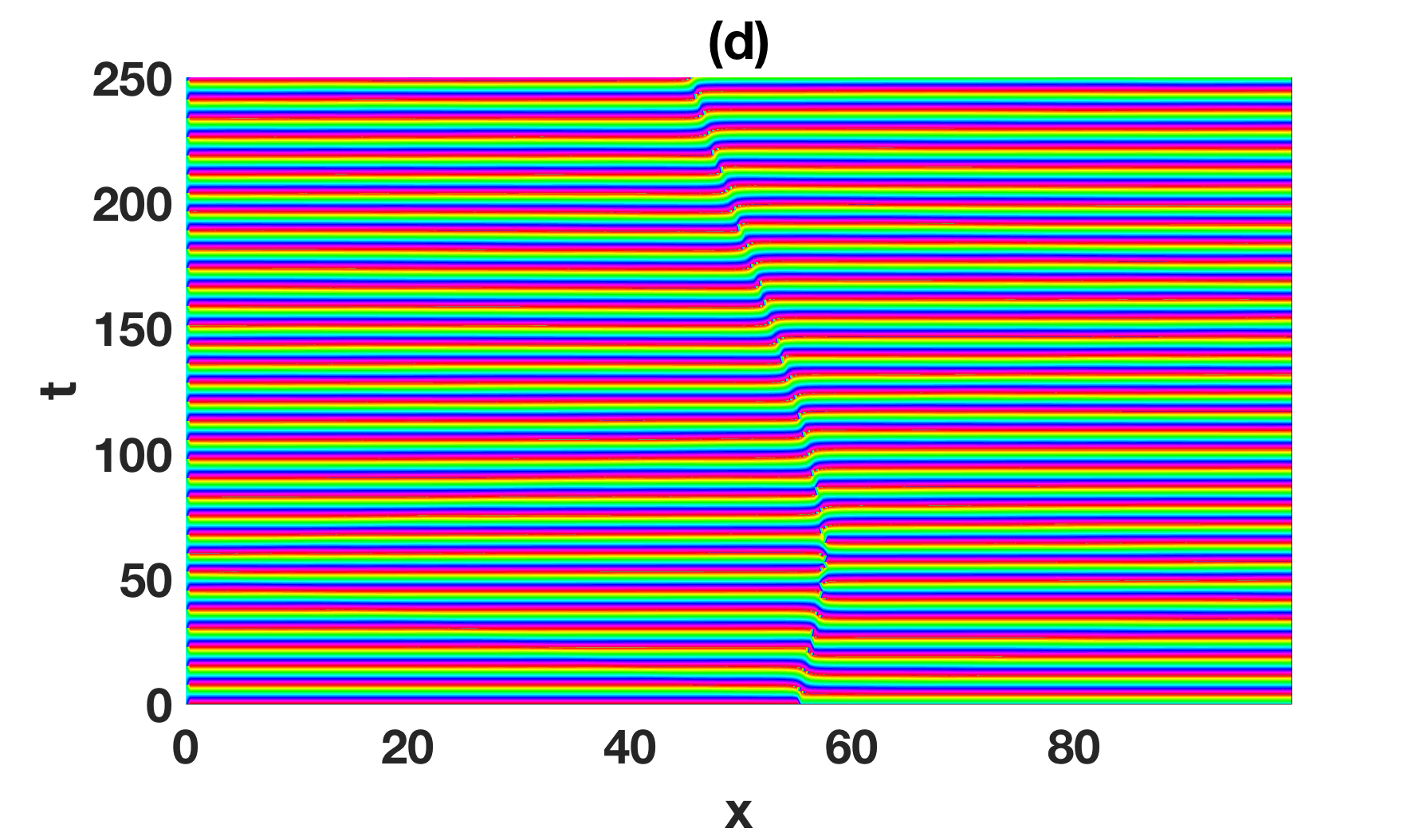}
  %\caption{1b}
  %\label{fig:phi_6}
\end{subfigure}
\caption{\emph{Simple reflection of a dark-bright soliton.} (a) Density and (b) phase of the bright soliton; (c) density and (d) phase of the dark soliton. The kinetic energy of the two components dark-bright soliton is less than the potential energy of the barrier and therefore the dark-bright soliton reflects from the barrier. We set  $\alpha = 0.15$ and $V_{\mathrm{CM}} = 0.06$. The delta function located at $x=60$. The white thick line represents the analytical results.} 
  \label{fig:FRHPRA:all_fig_delta_0-15}
\end{figure}

\begin{figure}
  \centering
  \includegraphics[width=\columnwidth]{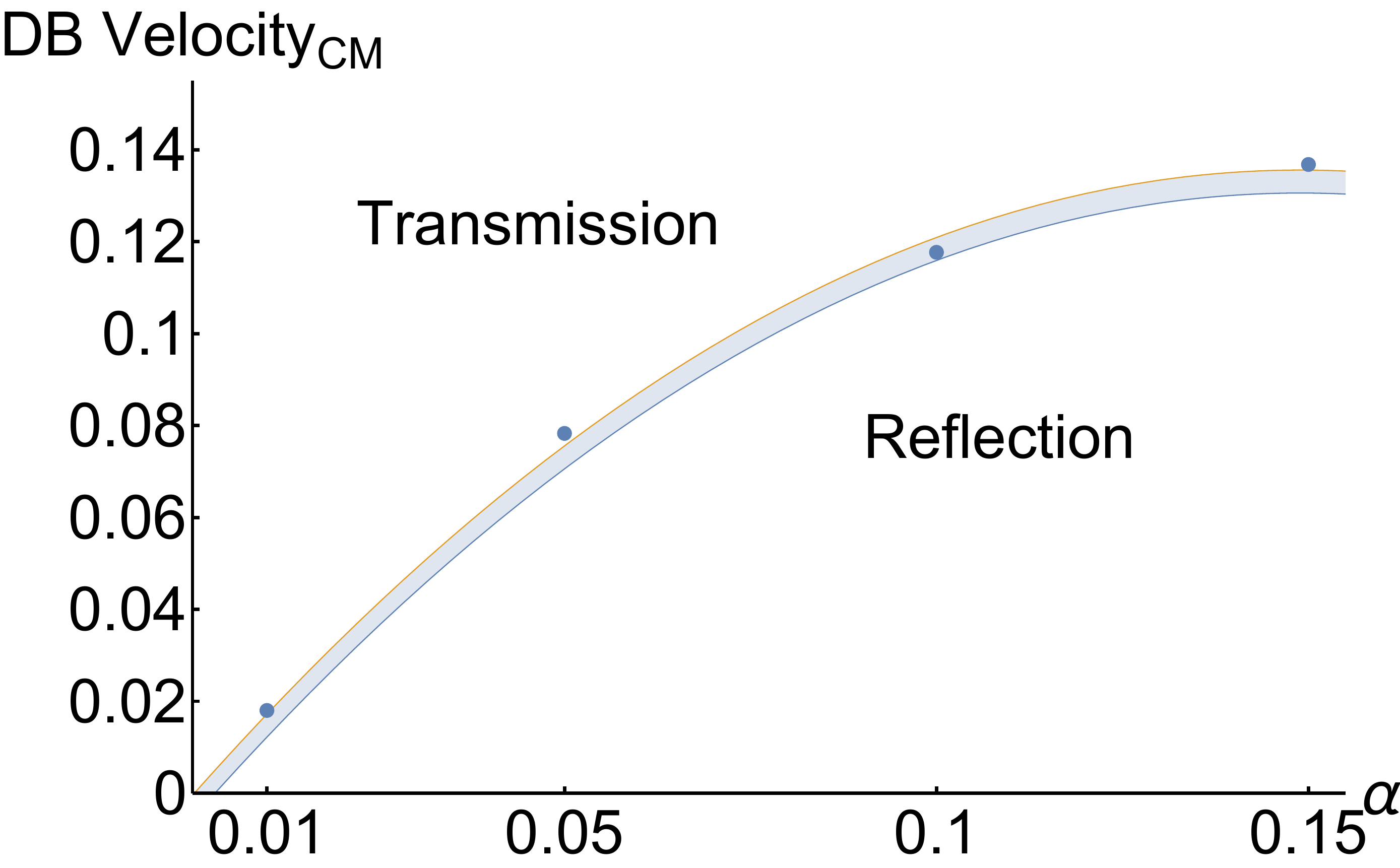}
  \caption{\emph{Transmission and reflection of dark-bright soliton for different values of the potential strength and center of mass velocity.} We compare the analytical predictions to numerical results for a wide range of the delta function amplitude, $\alpha$, and the dark-bright center of mass velocity, $V_{\mathrm{CM}}$. We identify the regions for the transmission and reflection of the dark-bright soliton by the potential barrier based on the parameter domain, $\alpha$ and $V_{\mathrm{CM}}$. The gray area represents inelastic scattering (i.e., internal excitation), showing that excitation of inelastic modes generally occur when close to the border between transmission and reflection. Note for $V_{\mathrm{CM}}=0.06$ we have a transmission of the dark-bright soliton for $\alpha=0.01$ and reflection when $\alpha=0.04$ and $0.15$ as described in Fig.~\ref{fig:FRHPRA:all_fig_delta_0-01}, \ref{fig:FRHPRA:all_fig_delta_0-04} and \ref{fig:FRHPRA:all_fig_delta_0-15}}
  \label{fig:FRHPRA:summary_plot}
\end{figure}

In contrast, in Fig.~\ref{fig:FRHPRA:all_fig_delta_0-15} we found that the dark-bright soliton reflects rapidly from the potential for $\alpha=0.15$. The delta function potential, in this case, does not allow for the creation of a quasibound state as in Fig.~\ref{fig:FRHPRA:all_fig_delta_0-15}. 

In Fig.~\ref{fig:FRHPRA:summary_plot}, we compare the analytical predictions to the numerical calculations for a wide range of delta function strength (i.e., $\alpha$) and the center of mass velocity of the dark-bright soliton. We identify three regions. The transmission of the dark-bright soliton over the barrier, the reflection, and the inelastic scattering region. These three case studies outline the basic kinds of dynamical outcomes. The dark-bright soliton has an additional characteristic that during the scattering process, for a small range of delta function strength, energy can be absorbed into the internal mode. In this case, the oscillation mode. We defined this region as an inelastic scattering region. It is noteworthy to mention that the inelastic scattering and the excitation of the internal modes occur only when we allow for an additional degree of freedom, as we do in this article, namely, the internal oscillation of the two components.

The basic idea is the scattering process interaction with the impurity transfers kinetic center of mass energy into internal modes, resulting in inelastic scattering.  Two of these modes are captured by the analytical model: the dominant feature of relative oscillation between the two components, as well as the oscillation in the widths.  However, the analytical model requires these widths oscillate in sync.  The numerical simulations allow further internal modes to enter the problem, starting with out-of-sync oscillations of the soliton widths, and including even shape deformations of various kinds. In general, the scattering of a dark-bright soliton is a complex inelastic process which will require experiments to properly understand, especially since quantum fluctuations are well known to concentrate at mean field minima, in this case the interstices where the bright soliton meets the dark soliton.  A proper treatment of such quantum fluctuations is an excellent subject for future study and involves at a minimum solution of the dynamical Bogoliubov equations. 
\begin{figure}
  \centering
  \includegraphics[width=\columnwidth]{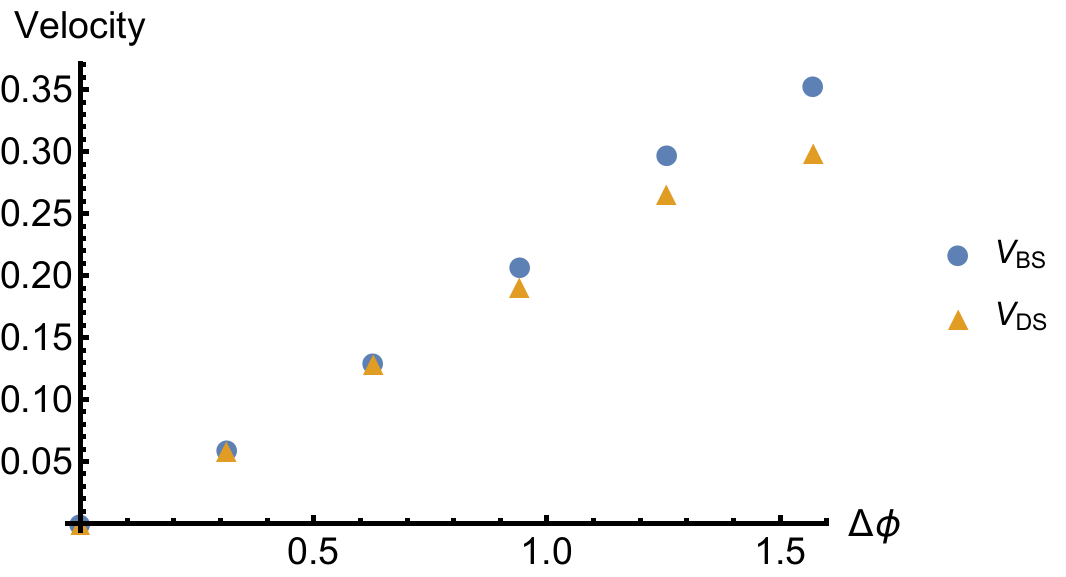}
  \caption{\emph{Dark-bright soliton component velocities.} We plot the velocities of the two components vs the phase difference imprinted on the dark component only, $\Delta \phi$. For a bright soliton component with half the amplitude of the dark soliton component the maximum velocity of the dark-bright soliton before it oscillates is half the speed of sound, $c/2$, as predicted from Fig.~\ref{fig:FRHPRA:fig1}. In the simulation units, $c/2=0.15$. Above this value, the two components start to oscillate.}
  \label{fig:FRHPRA:DB_velocity_num}
\end{figure}
\subsection{Dark-bright soliton velocity}
\label{sec:FRHPRA:Numerical_Dark-bright soliton velocity}
The behavior of the dark soliton velocity changes when interacting with another component, in this case, a bright soliton component in a dark-bright soliton. To study this behavior numerically, we imprint a phase jump, $\Delta \phi$,  on the dark soliton component only. In this way, we adiabatically move the two components such that we do not cause an oscillation between them, to explore our analytical predictions for the Manakov case from Sec.~\ref{sec:VS_velocity}. It is important to mention that the interaction coefficients (i.e., $g_{1}$ and $g_{2}$) are all positive in this case. This means that the bright soliton component can only live in such repulsive media by interacting with the dark soliton component. As mentioned in Sec.~\ref{sec:VS_velocity}, Eq.~\eqref{eq:coupledNLSE_VS} possesses an exact analytical dark-bright soliton solution, Eq.~\eqref{eq:DB_solution}. By examining this solution, we find that both component locations of the dark-bright soliton are expressed by one single spatial variable, $x_{0}\left(t\right)$. This is a criterion for an exact solution of Eq.~\eqref{eq:coupledNLSE_VS}.

In Fig.~\ref{fig:FRHPRA:fig1}, we see that the existence of the bright soliton component with half the amplitude of the dark soliton component prevents the dark soliton component from reaching its maximum velocity, $\mu_{1}$ and puts an upper limit on it. This is the upper limit for the velocity of the dark-bright soliton before the two components oscillate. By adopting the method mentioned above to move the dark-bright soliton we are in a position to compare the analytical results obtained in Sec.~\ref{sec:VS_velocity} with the numerical results we have in this section.

In Fig.~\ref{fig:FRHPRA:DB_velocity_num}, we imprint a phase difference on the dark soliton component only with interaction parameters $g_{1}=2$, $g_{2}=3$ and $g=2.6$. We find that the two components in the dark-bright soliton have the same velocity below a critical value of the phase imprinted. Therefore, no internal oscillation happens and the one variable,$x_{0}\left(t\right)$, represents the two-component locations. Above the critical value, we find the two components start to acquire different velocities. Consequently, an internal oscillation between the two components occurs and the positions of the dark component and the bright component no longer coincide. Therefore, the two-component cannot be expressed by one variable as described in Eq.~\eqref{eq:DB_solution}. 
\section{Conclusions}
	\label{sec:FRHPRA:Conclusions}
We obtained a system of equation of motions for a dark-bright soliton scattering off a fixed localized impurity, modeled by a delta function potential. We used a variational method with a hyperbolic tangent for the dark component and a hyperbolic secant for the bright component. The existence of the delta function altered the background of the dark soliton component, and therefore a perturbation method was needed to incorporate the effect of the delta function. The interaction of the dark-bright soliton with the potential excites different modes in the system. As a result, the dark-bright soliton emerges with a different velocity. Our analytical model capture two of these modes: the dominant feature of relative oscillation between the two components, as well as the oscillation in the widths. However, the analytical model requires these widths oscillate in sync. The numerical simulations allow further internal modes to enter the problem, starting with out-of-sync oscillations of the soliton widths, and including even shape deformations of various kinds. 

We identify regions for the transmission, reflection and inelastic scattering of the dark-bright soliton by the potential barrier. We present three case studies outlining the basic kinds of dynamical outcomes. The many internal modes excited in this problem show the complexity of the nonlinear dynamical multicomponent problem. Our study rather points to different physical regimes, and one can follow up by applying our model to any particular experiment intending to pursue the scattering question.  Nevertheless, we have provided at least one case study of transmission/reflection in Fig.~\ref{fig:FRHPRA:summary_plot}, to give the reader a general idea of the sorts of regimes that may occur. The scattering of a dark-bright soliton could also cause quantum fluctuations, as one might model, e.g., in dynamical Bogoliubov theory. In this case, the kinetic energy would go not only into internal mean-field modes but also into enhanced quantum fluctuations localized in and near the dark-bright soliton.  If that is the case, then a reduced velocity of a scattered dark-bright soliton beyond mean-field predictions will be a sign of quantum fluctuations.  This is another strong reason to get the mean-field inelastic scattering correct, carefully understanding all internal modes created by interaction with the impurity.    

In scattering theory, we usually put no limit on the incident kinetic energy.  However, dark solitons are well known to be limited to the speed of sound c in the medium.  The dark soliton grows shallower as the velocity is increased and eventually disappears.  The dark-bright soliton is also limited in velocity and therefore incident kinetic energy.  However, the limit is much more stringent.  We showed in the Manakov or equal-interaction case where it scales with the relative number of atoms in the bright and dark components. That is, as the dark soliton goes faster and is therefore shallower, it can no longer support the bright soliton. For example, when the bright soliton has half of the number of atoms as the dark one excavates or pushes aside, the maximum velocity is half the sound speed. Above this critical velocity the soliton components begin to oscillate, and eventually break apart.  This limits the kind of scattering experiments that can be performed in multicomponent BEC experiments and presents a smoking gun signal.

Future work may extend the investigation of the interaction of vector soliton with an impurity to three-component. We might rip apart the dark-bright soliton with the proper resonance condition, as found for exciton transport. In this sense, the barrier can be used to reflect, transmit, excite, or even destroy the dark-bright soliton~\cite{Zang2017,Lusk2015}. In addition, by solving this single impurity problem, we may extend the work for solving the disordered problem. It is noteworthy to mention that the excitation of the internal modes occur only when we allow for an additional degree of freedom, as we do in this article, namely, the internal oscillation of the two components which reflect the importance of using ansatz with two independent positions for the dark and bright soliton components.   
%
%%%%%%%%%%%%%%%%%%%%%%%%%%%%%   acknowledgments   %%%%%%%%%%%%%%%%%%%%%%%%%%%%%%
\acknowledgments
This material is based in part upon work supported by the US National Science Foundation under grant numbers PHY-1520915, OAC-1740130, as well as the US Air Force Office of Scientific Research grant number FA9550-14-1-0287. 

% We acknowledge support of the U.K. Engineering and Physical Sciences Research Council (EPSRC) through the “Quantum Science with Ultracold Molecules” Programme (Grant No. EP/P01058X/1)
   
%%%%%%%%%%%%%%%%%%%%%%%%%%%%%%%%%%%%%%%%%%%%%%%%%%%%%%%%%%%%%%%%%%%%%%%%%%%%%%%%%%%%
%
\FloatBarrier
\bibliographystyle{apsrev4-1}
\bibliography{library}

\begin{thebibliography}{26}%
\makeatletter
\providecommand \@ifxundefined [1]{%
 \@ifx{#1\undefined}
}%
\providecommand \@ifnum [1]{%
 \ifnum #1\expandafter \@firstoftwo
 \else \expandafter \@secondoftwo
 \fi
}%
\providecommand \@ifx [1]{%
 \ifx #1\expandafter \@firstoftwo
 \else \expandafter \@secondoftwo
 \fi
}%
\providecommand \natexlab [1]{#1}%
\providecommand \enquote  [1]{``#1''}%
\providecommand \bibnamefont  [1]{#1}%
\providecommand \bibfnamefont [1]{#1}%
\providecommand \citenamefont [1]{#1}%
\providecommand \href@noop [0]{\@secondoftwo}%
\providecommand \href [0]{\begingroup \@sanitize@url \@href}%
\providecommand \@href[1]{\@@startlink{#1}\@@href}%
\providecommand \@@href[1]{\endgroup#1\@@endlink}%
\providecommand \@sanitize@url [0]{\catcode `\\12\catcode `\$12\catcode
  `\&12\catcode `\#12\catcode `\^12\catcode `\_12\catcode `\%12\relax}%
\providecommand \@@startlink[1]{}%
\providecommand \@@endlink[0]{}%
\providecommand \url  [0]{\begingroup\@sanitize@url \@url }%
\providecommand \@url [1]{\endgroup\@href {#1}{\urlprefix }}%
\providecommand \urlprefix  [0]{URL }%
\providecommand \Eprint [0]{\href }%
\providecommand \doibase [0]{http://dx.doi.org/}%
\providecommand \selectlanguage [0]{\@gobble}%
\providecommand \bibinfo  [0]{\@secondoftwo}%
\providecommand \bibfield  [0]{\@secondoftwo}%
\providecommand \translation [1]{[#1]}%
\providecommand \BibitemOpen [0]{}%
\providecommand \bibitemStop [0]{}%
\providecommand \bibitemNoStop [0]{.\EOS\space}%
\providecommand \EOS [0]{\spacefactor3000\relax}%
\providecommand \BibitemShut  [1]{\csname bibitem#1\endcsname}%
\let\auto@bib@innerbib\@empty
%</preamble>
\bibitem [{\citenamefont {Griffiths}(2005)}]{griffiths2005introduction}%
  \BibitemOpen
  \bibfield  {author} {\bibinfo {author} {\bibfnamefont {D.~J.}\ \bibnamefont
  {Griffiths}},\ }\href {https://books.google.com/books?id=z4fwAAAAMAAJ} {\emph
  {\bibinfo {title} {{Introduction to Quantum Mechanics}}}},\ Pearson
  international edition\ (\bibinfo  {publisher} {Pearson Prentice Hall},\
  \bibinfo {year} {2005})\BibitemShut {NoStop}%
\bibitem [{\citenamefont {Kevrekidis}\ \emph {et~al.}(2008)\citenamefont
  {Kevrekidis}, \citenamefont {Frantzeskakis}, \citenamefont
  {Carretero-Gonz{\'{a}}lez}, \citenamefont {Parker}, \citenamefont {Jackson},
  \citenamefont {Martin},\ and\ \citenamefont {Adams}}]{Kevrekidis2008}%
  \BibitemOpen
  \bibfield  {author} {\bibinfo {author} {\bibfnamefont {P.~G.}\ \bibnamefont
  {Kevrekidis}}, \bibinfo {author} {\bibfnamefont {D.~J.}\ \bibnamefont
  {Frantzeskakis}}, \bibinfo {author} {\bibfnamefont {R.}~\bibnamefont
  {Carretero-Gonz{\'{a}}lez}}, \bibinfo {author} {\bibfnamefont {N.~G.}\
  \bibnamefont {Parker}}, \bibinfo {author} {\bibfnamefont {B.}~\bibnamefont
  {Jackson}}, \bibinfo {author} {\bibfnamefont {A.~M.}\ \bibnamefont {Martin}},
  \ and\ \bibinfo {author} {\bibfnamefont {C.~S.}\ \bibnamefont {Adams}},\
  }\href {\doibase 10.1007/978-3-540-73591-5} {\emph {\bibinfo {title}
  {{Emergent Nonlinear Phenomena in Bose-Einstein Condensates}}}}\ (\bibinfo
  {publisher} {Springer Series},\ \bibinfo {year} {2008})\BibitemShut {NoStop}%
\bibitem [{\citenamefont {Kivshar}\ and\ \citenamefont
  {Malomed}(1989)}]{Kivshar1989b}%
  \BibitemOpen
  \bibfield  {author} {\bibinfo {author} {\bibfnamefont {Y.}~\bibnamefont
  {Kivshar}}\ and\ \bibinfo {author} {\bibfnamefont {B.}~\bibnamefont
  {Malomed}},\ }\href {http://rmp.aps.org/abstract/RMP/v61/i4/p763_1}
  {\bibfield  {journal} {\bibinfo  {journal} {Reviews of Modern Physics}\
  }\textbf {\bibinfo {volume} {61}} (\bibinfo {year} {1989})}\BibitemShut
  {NoStop}%
\bibitem [{\citenamefont {Kosevich}(1990)}]{Kosevich1990}%
  \BibitemOpen
  \bibfield  {author} {\bibinfo {author} {\bibfnamefont {A.~M.}\ \bibnamefont
  {Kosevich}},\ }\href {\doibase 10.1016/0167-2789(90)90126-A} {\bibfield
  {journal} {\bibinfo  {journal} {Physica D: Nonlinear Phenomena}\ }\textbf
  {\bibinfo {volume} {41}},\ \bibinfo {pages} {253} (\bibinfo {year}
  {1990})}\BibitemShut {NoStop}%
\bibitem [{\citenamefont {Lee}\ and\ \citenamefont {Brand}(2006)}]{Lee2006}%
  \BibitemOpen
  \bibfield  {author} {\bibinfo {author} {\bibfnamefont {C.}~\bibnamefont
  {Lee}}\ and\ \bibinfo {author} {\bibfnamefont {J.}~\bibnamefont {Brand}},\
  }\href {\doibase 10.1209/epl/i2005-10408-4} {\bibfield  {journal} {\bibinfo
  {journal} {Europhysics Letters (EPL)}\ }\textbf {\bibinfo {volume} {73}},\
  \bibinfo {pages} {321} (\bibinfo {year} {2006})}\BibitemShut {NoStop}%
\bibitem [{\citenamefont {Frantzeskakis}\ \emph {et~al.}(2002)\citenamefont
  {Frantzeskakis}, \citenamefont {Theocharis}, \citenamefont {Diakonos},
  \citenamefont {Schmelcher},\ and\ \citenamefont
  {Kivshar}}]{Frantzeskakis2002}%
  \BibitemOpen
  \bibfield  {author} {\bibinfo {author} {\bibfnamefont {D.}~\bibnamefont
  {Frantzeskakis}}, \bibinfo {author} {\bibfnamefont {G.}~\bibnamefont
  {Theocharis}}, \bibinfo {author} {\bibfnamefont {F.}~\bibnamefont
  {Diakonos}}, \bibinfo {author} {\bibfnamefont {P.}~\bibnamefont
  {Schmelcher}}, \ and\ \bibinfo {author} {\bibfnamefont {Y.}~\bibnamefont
  {Kivshar}},\ }\href {\doibase 10.1103/PhysRevA.66.053608} {\bibfield
  {journal} {\bibinfo  {journal} {Physical Review A}\ }\textbf {\bibinfo
  {volume} {66}},\ \bibinfo {pages} {053608} (\bibinfo {year}
  {2002})}\BibitemShut {NoStop}%
\bibitem [{\citenamefont {Garnier}(2006)}]{Garnier2006}%
  \BibitemOpen
  \bibfield  {author} {\bibinfo {author} {\bibfnamefont {J.}~\bibnamefont
  {Garnier}},\ }\href {\doibase 10.1103/PhysRevA.74.013604} {\bibfield
  {journal} {\bibinfo  {journal} {Physical Review A}\ }\textbf {\bibinfo
  {volume} {74}},\ \bibinfo {pages} {013604} (\bibinfo {year}
  {2006})}\BibitemShut {NoStop}%
\bibitem [{\citenamefont {Seaman}\ \emph {et~al.}(2005)\citenamefont {Seaman},
  \citenamefont {Carr},\ and\ \citenamefont {Holland}}]{Seaman2005a}%
  \BibitemOpen
  \bibfield  {author} {\bibinfo {author} {\bibfnamefont {B.}~\bibnamefont
  {Seaman}}, \bibinfo {author} {\bibfnamefont {L.}~\bibnamefont {Carr}}, \ and\
  \bibinfo {author} {\bibfnamefont {M.}~\bibnamefont {Holland}},\ }\href
  {\doibase 10.1103/PhysRevA.71.033609} {\bibfield  {journal} {\bibinfo
  {journal} {Physical Review A}\ }\textbf {\bibinfo {volume} {71}},\ \bibinfo
  {pages} {033609} (\bibinfo {year} {2005})}\BibitemShut {NoStop}%
\bibitem [{\citenamefont {Sykes}\ \emph {et~al.}(2009)\citenamefont {Sykes},
  \citenamefont {Davis},\ and\ \citenamefont {Roberts}}]{Sykes2009}%
  \BibitemOpen
  \bibfield  {author} {\bibinfo {author} {\bibfnamefont {A.~G.}\ \bibnamefont
  {Sykes}}, \bibinfo {author} {\bibfnamefont {M.~J.}\ \bibnamefont {Davis}}, \
  and\ \bibinfo {author} {\bibfnamefont {D.~C.}\ \bibnamefont {Roberts}},\
  }\href {\doibase 10.1103/PhysRevLett.103.085302} {\bibfield  {journal}
  {\bibinfo  {journal} {Physical Review Letters}\ }\textbf {\bibinfo {volume}
  {103}},\ \bibinfo {pages} {085302} (\bibinfo {year} {2009})}\BibitemShut
  {NoStop}%
\bibitem [{\citenamefont {Achilleos}\ \emph {et~al.}(2011)\citenamefont
  {Achilleos}, \citenamefont {Kevrekidis},\ and\ \citenamefont
  {Rothos}}]{Achilleos2011}%
  \BibitemOpen
  \bibfield  {author} {\bibinfo {author} {\bibfnamefont {V.}~\bibnamefont
  {Achilleos}}, \bibinfo {author} {\bibfnamefont {P.}~\bibnamefont
  {Kevrekidis}}, \ and\ \bibinfo {author} {\bibfnamefont {V.}~\bibnamefont
  {Rothos}},\ }\href {\doibase 10.1103/PhysRevA.84.053626} {\bibfield
  {journal} {\bibinfo  {journal} {Physical Review A}\ ,\ \bibinfo {pages}
  {053626}} (\bibinfo {year} {2011})}\BibitemShut {NoStop}%
\bibitem [{\citenamefont {Sykes}(2011)}]{Sykes2011}%
  \BibitemOpen
  \bibfield  {author} {\bibinfo {author} {\bibfnamefont {A.}~\bibnamefont
  {Sykes}},\ }\href {\doibase 10.1088/1751-8113/44/13/135206} {\bibfield
  {journal} {\bibinfo  {journal} {Journal of Physics A Mathematical and
  General}\ }\textbf {\bibinfo {volume} {44}},\ \bibinfo {pages} {135206}
  (\bibinfo {year} {2011})}\BibitemShut {NoStop}%
\bibitem [{\citenamefont {Hoefer}\ \emph {et~al.}(2011)\citenamefont {Hoefer},
  \citenamefont {Chang}, \citenamefont {Hamner},\ and\ \citenamefont
  {Engels}}]{Hoefer2011b}%
  \BibitemOpen
  \bibfield  {author} {\bibinfo {author} {\bibfnamefont {M.~A.}\ \bibnamefont
  {Hoefer}}, \bibinfo {author} {\bibfnamefont {J.~J.}\ \bibnamefont {Chang}},
  \bibinfo {author} {\bibfnamefont {C.}~\bibnamefont {Hamner}}, \ and\ \bibinfo
  {author} {\bibfnamefont {P.}~\bibnamefont {Engels}},\ }\href
  {http://link.aps.org/doi/10.1103/PhysRevA.84.041605} {\bibfield  {journal}
  {\bibinfo  {journal} {Physical Review A}\ }\textbf {\bibinfo {volume} {84}},\
  \bibinfo {pages} {41605} (\bibinfo {year} {2011})}\BibitemShut {NoStop}%
\bibitem [{\citenamefont {Yan}\ \emph {et~al.}(2012)\citenamefont {Yan},
  \citenamefont {Chang}, \citenamefont {Hamner}, \citenamefont {Hoefer},
  \citenamefont {Kevrekidis}, \citenamefont {Engels}, \citenamefont
  {Achilleos}, \citenamefont {Frantzeskakis},\ and\ \citenamefont
  {Cuevas}}]{Yan2012}%
  \BibitemOpen
  \bibfield  {author} {\bibinfo {author} {\bibfnamefont {D.}~\bibnamefont
  {Yan}}, \bibinfo {author} {\bibfnamefont {J.~J.}\ \bibnamefont {Chang}},
  \bibinfo {author} {\bibfnamefont {C.}~\bibnamefont {Hamner}}, \bibinfo
  {author} {\bibfnamefont {M.}~\bibnamefont {Hoefer}}, \bibinfo {author}
  {\bibfnamefont {P.~G.}\ \bibnamefont {Kevrekidis}}, \bibinfo {author}
  {\bibfnamefont {P.}~\bibnamefont {Engels}}, \bibinfo {author} {\bibfnamefont
  {V.}~\bibnamefont {Achilleos}}, \bibinfo {author} {\bibfnamefont {D.~J.}\
  \bibnamefont {Frantzeskakis}}, \ and\ \bibinfo {author} {\bibfnamefont
  {J.}~\bibnamefont {Cuevas}},\ }\href
  {http://stacks.iop.org/0953-4075/45/i=11/a=115301} {\bibfield  {journal}
  {\bibinfo  {journal} {Journal of Physics B: Atomic, Molecular and Optical
  Physics}\ }\textbf {\bibinfo {volume} {45}},\ \bibinfo {pages} {115301}
  (\bibinfo {year} {2012})}\BibitemShut {NoStop}%
\bibitem [{\citenamefont {Hamner}\ \emph {et~al.}(2011)\citenamefont {Hamner},
  \citenamefont {Chang}, \citenamefont {Engels},\ and\ \citenamefont
  {Hoefer}}]{Hamner2011b}%
  \BibitemOpen
  \bibfield  {author} {\bibinfo {author} {\bibfnamefont {C.}~\bibnamefont
  {Hamner}}, \bibinfo {author} {\bibfnamefont {J.~J.}\ \bibnamefont {Chang}},
  \bibinfo {author} {\bibfnamefont {P.}~\bibnamefont {Engels}}, \ and\ \bibinfo
  {author} {\bibfnamefont {M.~A.}\ \bibnamefont {Hoefer}},\ }\href
  {http://link.aps.org/doi/10.1103/PhysRevLett.106.065302} {\bibfield
  {journal} {\bibinfo  {journal} {Physical Review Letters}\ }\textbf {\bibinfo
  {volume} {106}},\ \bibinfo {pages} {65302} (\bibinfo {year}
  {2011})}\BibitemShut {NoStop}%
\bibitem [{\citenamefont {Alotaibi}\ and\ \citenamefont
  {Carr}(2017)}]{majed2017}%
  \BibitemOpen
  \bibfield  {author} {\bibinfo {author} {\bibfnamefont {M.~O.~D.}\
  \bibnamefont {Alotaibi}}\ and\ \bibinfo {author} {\bibfnamefont {L.~D.}\
  \bibnamefont {Carr}},\ }\href {\doibase 10.1103/PhysRevA.96.013601}
  {\bibfield  {journal} {\bibinfo  {journal} {Phys. Rev. A}\ }\textbf {\bibinfo
  {volume} {96}},\ \bibinfo {pages} {13601} (\bibinfo {year}
  {2017})}\BibitemShut {NoStop}%
\bibitem [{\citenamefont {Achilleos}\ \emph {et~al.}(2012)\citenamefont
  {Achilleos}, \citenamefont {Yan}, \citenamefont {Kevrekidis},\ and\
  \citenamefont {Frantzeskakis}}]{Frantzeskakis2012}%
  \BibitemOpen
  \bibfield  {author} {\bibinfo {author} {\bibfnamefont {V.}~\bibnamefont
  {Achilleos}}, \bibinfo {author} {\bibfnamefont {D.}~\bibnamefont {Yan}},
  \bibinfo {author} {\bibfnamefont {P.~G.}\ \bibnamefont {Kevrekidis}}, \ and\
  \bibinfo {author} {\bibfnamefont {D.~J.}\ \bibnamefont {Frantzeskakis}},\
  }\href {\doibase 10.1088/1367-2630/14/5/055006} {\bibfield  {journal}
  {\bibinfo  {journal} {New Journal of Physics}\ }\textbf {\bibinfo {volume}
  {14}},\ \bibinfo {pages} {55006} (\bibinfo {year} {2012})}\BibitemShut
  {NoStop}%
\bibitem [{\citenamefont {Kivshar}\ and\ \citenamefont
  {Yang}(1994)}]{Kivshar1994}%
  \BibitemOpen
  \bibfield  {author} {\bibinfo {author} {\bibfnamefont {Y.}~\bibnamefont
  {Kivshar}}\ and\ \bibinfo {author} {\bibfnamefont {X.}~\bibnamefont {Yang}},\
  }\href {http://www.sciencedirect.com/science/article/pii/0960077994901082}
  {\bibfield  {journal} {\bibinfo  {journal} {Physical Review E}\ }\textbf
  {\bibinfo {volume} {49}},\ \bibinfo {pages} {1657} (\bibinfo {year}
  {1994})}\BibitemShut {NoStop}%
\bibitem [{\citenamefont {Kivshar}\ and\ \citenamefont
  {Kr{\'{o}}likowski}(1995)}]{Kivshar1995}%
  \BibitemOpen
  \bibfield  {author} {\bibinfo {author} {\bibfnamefont {Y.}~\bibnamefont
  {Kivshar}}\ and\ \bibinfo {author} {\bibfnamefont {W.}~\bibnamefont
  {Kr{\'{o}}likowski}},\ }\href
  {http://www.sciencedirect.com/science/article/pii/003040189400644A}
  {\bibfield  {journal} {\bibinfo  {journal} {Optics communications}\ }\textbf
  {\bibinfo {volume} {114}},\ \bibinfo {pages} {353 } (\bibinfo {year}
  {1995})}\BibitemShut {NoStop}%
\bibitem [{\citenamefont {Frantzeskakis}(2010)}]{Frantzeskakis2010a}%
  \BibitemOpen
  \bibfield  {author} {\bibinfo {author} {\bibfnamefont {D.~J.}\ \bibnamefont
  {Frantzeskakis}},\ }\href {http://stacks.iop.org/1751-8121/43/i=21/a=213001}
  {\bibfield  {journal} {\bibinfo  {journal} {Journal of Physics A:
  Mathematical and Theoretical}\ }\textbf {\bibinfo {volume} {43}},\ \bibinfo
  {pages} {213001} (\bibinfo {year} {2010})}\BibitemShut {NoStop}%
\bibitem [{\citenamefont {Carretero-Gonz{\'{a}}lez}\ \emph
  {et~al.}(2008)\citenamefont {Carretero-Gonz{\'{a}}lez}, \citenamefont
  {Frantzeskakis},\ and\ \citenamefont {Kevrekidis}}]{Carretero-Gonzalez2008b}%
  \BibitemOpen
  \bibfield  {author} {\bibinfo {author} {\bibfnamefont {R.}~\bibnamefont
  {Carretero-Gonz{\'{a}}lez}}, \bibinfo {author} {\bibfnamefont {D.~J.}\
  \bibnamefont {Frantzeskakis}}, \ and\ \bibinfo {author} {\bibfnamefont
  {P.~G.}\ \bibnamefont {Kevrekidis}},\ }\href
  {http://stacks.iop.org/0951-7715/21/i=7/a=R01} {\bibfield  {journal}
  {\bibinfo  {journal} {Nonlinearity}\ }\textbf {\bibinfo {volume} {21}},\
  \bibinfo {pages} {R139} (\bibinfo {year} {2008})}\BibitemShut {NoStop}%
\bibitem [{\citenamefont {Kevrekidis}\ \emph {et~al.}(2015)\citenamefont
  {Kevrekidis}, \citenamefont {Frantzeskakis},\ and\ \citenamefont
  {Carretero-Gonz{\'{a}}lez}}]{Kevrekidis2015}%
  \BibitemOpen
  \bibfield  {author} {\bibinfo {author} {\bibfnamefont {P.~G.}\ \bibnamefont
  {Kevrekidis}}, \bibinfo {author} {\bibfnamefont {D.~J.}\ \bibnamefont
  {Frantzeskakis}}, \ and\ \bibinfo {author} {\bibfnamefont {R.}~\bibnamefont
  {Carretero-Gonz{\'{a}}lez}},\ }\href
  {https://books.google.com/books?id=1BVoCgAAQBAJ} {\emph {\bibinfo {title}
  {{The Defocusing Nonlinear Schrodinger Equation: From Dark Solitons to
  Vortices and Vortex Rings}}}}\ (\bibinfo  {publisher} {Society for Industrial
  and Applied Mathematics},\ \bibinfo {year} {2015})\BibitemShut {NoStop}%
\bibitem [{\citenamefont {Salasnich}\ and\ \citenamefont
  {Malomed}(2006)}]{Salasnich2006}%
  \BibitemOpen
  \bibfield  {author} {\bibinfo {author} {\bibfnamefont {L.}~\bibnamefont
  {Salasnich}}\ and\ \bibinfo {author} {\bibfnamefont {B.}~\bibnamefont
  {Malomed}},\ }\href {\doibase 10.1103/PhysRevA.74.053610} {\bibfield
  {journal} {\bibinfo  {journal} {Physical Review A}\ }\textbf {\bibinfo
  {volume} {74}},\ \bibinfo {pages} {053610} (\bibinfo {year}
  {2006})}\BibitemShut {NoStop}%
\bibitem [{\citenamefont {{\'{A}}lvarez}\ \emph {et~al.}(2013)\citenamefont
  {{\'{A}}lvarez}, \citenamefont {Cuevas}, \citenamefont {Romero},
  \citenamefont {Hamner}, \citenamefont {Chang}, \citenamefont {Engels},
  \citenamefont {Kevrekidis},\ and\ \citenamefont
  {Frantzeskakis}}]{Alvarez2013a}%
  \BibitemOpen
  \bibfield  {author} {\bibinfo {author} {\bibfnamefont {A.}~\bibnamefont
  {{\'{A}}lvarez}}, \bibinfo {author} {\bibfnamefont {J.}~\bibnamefont
  {Cuevas}}, \bibinfo {author} {\bibfnamefont {F.~R.}\ \bibnamefont {Romero}},
  \bibinfo {author} {\bibfnamefont {C.}~\bibnamefont {Hamner}}, \bibinfo
  {author} {\bibfnamefont {J.~J.}\ \bibnamefont {Chang}}, \bibinfo {author}
  {\bibfnamefont {P.}~\bibnamefont {Engels}}, \bibinfo {author} {\bibfnamefont
  {P.~G.}\ \bibnamefont {Kevrekidis}}, \ and\ \bibinfo {author} {\bibfnamefont
  {D.~J.}\ \bibnamefont {Frantzeskakis}},\ }\href
  {http://stacks.iop.org/0953-4075/46/i=6/a=065302?key=crossref.8f7c46fb498dea8976993c1c00be94f2}
  {\bibfield  {journal} {\bibinfo  {journal} {Journal of Physics B: Atomic,
  Molecular and Optical Physics}\ }\textbf {\bibinfo {volume} {46}},\ \bibinfo
  {pages} {065302} (\bibinfo {year} {2013})}\BibitemShut {NoStop}%
\bibitem [{\citenamefont {Reinhardt}(1997)}]{Reinhardt1997b}%
  \BibitemOpen
  \bibfield  {author} {\bibinfo {author} {\bibfnamefont {W.}~\bibnamefont
  {Reinhardt}},\ }\href {http://iopscience.iop.org/0953-4075/30/22/001}
  {\bibfield  {journal} {\bibinfo  {journal} {of Physics B: Atomic, Molecular
  and}\ }\textbf {\bibinfo {volume} {785}},\ \bibinfo {pages} {9} (\bibinfo
  {year} {1997})}\BibitemShut {NoStop}%
\bibitem [{\citenamefont {Zang}\ \emph {et~al.}(2017)\citenamefont {Zang},
  \citenamefont {Montangero}, \citenamefont {Carr},\ and\ \citenamefont
  {Lusk}}]{Zang2017}%
  \BibitemOpen
  \bibfield  {author} {\bibinfo {author} {\bibfnamefont {X.}~\bibnamefont
  {Zang}}, \bibinfo {author} {\bibfnamefont {S.}~\bibnamefont {Montangero}},
  \bibinfo {author} {\bibfnamefont {L.~D.}\ \bibnamefont {Carr}}, \ and\
  \bibinfo {author} {\bibfnamefont {M.~T.}\ \bibnamefont {Lusk}},\ }\href
  {\doibase 10.1103/PhysRevB.95.195423} {\bibfield  {journal} {\bibinfo
  {journal} {Physical Review B}\ }\textbf {\bibinfo {volume} {95}},\ \bibinfo
  {pages} {195423} (\bibinfo {year} {2017})}\BibitemShut {NoStop}%
\bibitem [{\citenamefont {Lusk}\ \emph {et~al.}(2015)\citenamefont {Lusk},
  \citenamefont {Stafford}, \citenamefont {Zimmerman},\ and\ \citenamefont
  {Carr}}]{Lusk2015}%
  \BibitemOpen
  \bibfield  {author} {\bibinfo {author} {\bibfnamefont {M.~T.}\ \bibnamefont
  {Lusk}}, \bibinfo {author} {\bibfnamefont {C.~A.}\ \bibnamefont {Stafford}},
  \bibinfo {author} {\bibfnamefont {J.~D.}\ \bibnamefont {Zimmerman}}, \ and\
  \bibinfo {author} {\bibfnamefont {L.~D.}\ \bibnamefont {Carr}},\ }\href
  {\doibase 10.1103/PhysRevB.92.241112} {\bibfield  {journal} {\bibinfo
  {journal} {Physical Review B}\ }\textbf {\bibinfo {volume} {92}},\ \bibinfo
  {pages} {241112} (\bibinfo {year} {2015})}\BibitemShut {NoStop}%
\end{thebibliography}%
% \bibliographystyle{prsty}
% \bibliography{/Users/majedalotaibi/Documents/library}
\end{document}